\documentclass[showpacs,printnumbers,amsmath,amssymb,twocolumn, aps, pra, longbibliography]{revtex4-1}

\usepackage{graphicx}
\usepackage{dcolumn}
\usepackage{bm}
\usepackage{xspace} 
\usepackage{color}
\usepackage{enumerate}

\newcommand{\ket}[1]{|{#1}\rangle}

\def\Rb87{$^{87}\text{Rb}$}
\def\0{\ket{0}}
\def\1{\ket{1}}

\begin{document}


\title{Robustifying Twist-and-Turn Entanglement with Interaction-Based Readout}

\author{Safoura S. Mirkhalaf}
\affiliation{QSTAR, Largo Enrico Fermi 2, 50125, Firenze, Italy}
\author{Samuel P. Nolan} 
\affiliation{School of Mathematics and Physics, The University of Queensland, Brisbane, Queensland, Australia}
\author{Simon A. Haine}
\affiliation{Department of Physics and Astronomy, University of Sussex, Brighton, United Kingdom}
\email{simon.a.haine@gmail.com}

\date{\today}

\begin{abstract}

\noindent The use of multi-particle entangled states has the potential to drastically increase the sensitivity of atom interferometers and atomic clocks. The twist-and-turn (TNT) Hamiltonian can create multi-particle entanglement much more rapidly than the ubiquitous one-axis twisting (OAT) Hamiltonian in the same spin system. In this paper, we consider the effects of detection noise - a key limitation in current experiments - on the metrological usefulness of nonclassical states generated under TNT dynamics. We also consider a variety of interaction-based readouts to maximize their performance. Interestingly, the optimum interaction-based readout is not the obvious case of perfect time reversal.

\end{abstract}

\maketitle

\section{Introduction}
\noindent There is currently considerable interest in methods that can produce highly entangled states of atomic ensembles with the motivation of enhancing the sensitivity of atom interferometers and atomic clocks \cite{Pezze:2016_review}. Without many-body entanglement, the phase sensitivity of such experiments is fundamentally shot-noise-limited (SNL) $\Delta \phi =1/\sqrt{N}$ \cite{Giovannetti:2006, Pezze:2009}. In recent years, experiments in atomic systems based on the one-axis twisting (OAT) spin squeezing scheme of Kitagawa and Ueda \cite{Kitagawa:1993, Molmer:1999} have demonstrated metrologically useful spin-squeezing \cite{Esteve:2008, Leroux:2010}, and sub-shot-noise phase detection \cite{Gross:2010, Riedel:2010, Muessel:2014}. However, typical experiments are limited to only moderate quantum enhancement due to constraints on the state preparation time imposed by dephasing \cite{Li:2008, Li:2009} and multi-mode dynamics \cite{Haine:2009, Haine:2014}. This leads to a degree of quantum enhancement that is considerably less than the theoretical optimum. Recently, a related method known as twist-and-turn (TNT) squeezing has been demonstrated \cite{Micheli:2003,Strobel:2014, Muessel:2015, Sorelli:2015}. The TNT Hamiltonian, which uses the same nonlinear interactions as OAT with an additional linear rotation, typically produces more quantum enhancement for the same interaction time. As TNT is based on the same interactions that leads to OAT squeezing, it can be implemented in the same experimental set-ups. 

In practice, however, it is very difficult to fully exploit the nonclassical features of quantum states. This is mainly due to the fragility of quantum correlated states under inevitable sources of noise (e.g.\ phase or detection noise). One way to conquer these effects is to use \emph{interaction-based readout} protocols \cite{Marino:2012, Hosten:2016, Davis:2016, Macri:2016, Frowis:2016, Szigeti:2017, Davis:2017, Nolan:2017b, Huang:2017,  Anders:2018, Fang:2017, Hayes:2018} which make use of the appropriate unitary/supplementary operations right before final measurement. These schemes can be summarised as follows. First there is a state preparation stage, where the quantum Fisher information (QFI, denoted $F_Q$) of the (typically un-entangled) initial state $|\psi_0\rangle$ is increased via the unitary operator $U_1$. The QFI bounds the precision of an estimate via the quantum Cram{\'e}r-Rao bound (QCRB), which states $\Delta \phi \geq1/\sqrt{F_Q}$. The classical parameter to be estimated $\phi$ is then encoded onto the state via a measurement device (for example, a Mach-Zehnder interferometer) described by the unitary operator $U_\phi$. The interaction-based readout is then implemented by applying another unitary operator $\hat{U}_2$ directly before the final measurement, such that the final state is
\begin{eqnarray}
|\psi \rangle=U_2 U_{\phi} U_{1} |\psi_0 \rangle.
\end{eqnarray} 
$U_2$ does not alter the QFI, but \emph{can} effect the classical Fisher information (CFI, denoted $F_c$) when a measurement is made in a particular basis. Specifically, it has been shown that protocols which perfectly time reverse the initial unitary operator ($U_2=U_1^{\dagger}$, which we refer to as an \emph{echo}) followed by a measurement that projects onto the initial state, saturates the QCRB \cite{Macri:2016}, indicating the measurement is optimal. It has also been shown that using $U_2 = U_1^\dag$ can improve robustness against detection noise \cite{Davis:2016, Nolan:2017b}. However, in \cite{Nolan:2017b} it was shown that when a measurement that resolves the probability of \emph{all} results in a particular basis is made, there are \emph{many} choices of $U_2$ (including the trivial choice $U_2 = 1$) that saturate the QCRB. Furthermore, in many circumstances, there are choices for $U_2$ that provide greater robustness than $U_2 = U_1^\dag$. In this paper we investigate how interaction-based readouts improve robustness specifically for the case where the state preparation ($U_1$) and interaction based read-out ($U_2$) are based on the TNT interaction.

\section{Interaction-based readout protocol}\label{protocol}
\noindent Let us consider an ensemble of $N$ two-level atoms which are prepared in initial state $|\psi_0\rangle$.
The interaction-based readout protocol includes the following four steps: 

\begin{enumerate}[i]
\item \textit{State preparation:} This increases the QFI of the initial state via unitary evolution $U_1$ as 
$\rho=U_1 |\psi_0\rangle \langle\psi_0 | U_1^{\dagger} $. In particular, we use the TNT Hamiltonian for this purpose [defined in the next Section, Eq.~\eqref{H}]. 

\item \textit{Phase encoding:} An unknown phase is encoded into the state by performing linear unitary $U_{\phi}=e^{-i\phi \hat{S}_n}$ on the system such that $U_{\phi}^{\dagger}\rho U_{\phi}$. This process corresponds to making a rotation around an arbitrary spin direction $\vec{\bm{n}}$. 

\item \textit{Interaction-based readout:} A second entangling evolution $U_2$ is applied onto the state
as $U_2U_{\phi}\rho U_{\phi}^{\dagger} U_{2}^{\dagger}$. Echo protocols refer to the special case of perfect time reversal
$U_2=U_1^{\dagger}$. Under certain conditions, which we demonstrate below, $U_2$ allows measurements made in the appropriate basis to saturate the QCRB.

\item \textit{Spin-resolving measurement:} A spin-resolving measurement projects on to some complete orthogonal basis $\{ |m\rangle \}$. This is equivalent to estimating the full probability distribution $P_m(\phi)=\langle m|U_2U_{\phi}\rho U_{\phi}^{\dagger} U_{2}^{\dagger}|m \rangle$. 
\end{enumerate}

\noindent Consequently, the phase $\phi$ can be estimated by examining the probabilities 
$P_m(\phi)$. Based on parameter estimation theory, the precision of the estimate is bounded by $\Delta\phi^2=1/{F_c(\phi)}$ where
\begin{eqnarray}\label{FC}
F_c(\phi) = \sum_{m} \frac{1}{P_m(\phi)}\bigg(\frac{\partial P_m(\phi)}{\partial\phi}\bigg)^2
\end{eqnarray}
is the classical Fisher information (CFI). For single-variate parameter estimation, it can be shown that a measurement basis yielding $F_c = F_Q$ is guaranteed to exist \cite{Braunstein:1994, Toth:2014, Demkowicz-Dobrzanski:2014}. This measurement is optimal - the measurement saturates QCRB, which is the ultimate sensitivity limit imposed by $\rho$. As was shown in \cite{Nolan:2017b}, the conditions that $U_2$ and the measurement basis $\{|m\rangle \}$ saturate the QCRB are (see Appendix \ref{sec:appendixa} for a more detailed proof):
\begin{enumerate}
\item The initial state is the parity eigenstate in the measurement basis $\{ |m\rangle \}$; i.e.\ $\hat{\Pi} \rho= (-1)^p \rho$
where, $p=0,1$ for $\hat{\Pi}=\sum_m (-1)^m |m\rangle\langle m|$.

\item The generator $\hat{S}_n$ flips the parity; i.e.\ $\hat{\Pi} \hat{S}_n\hat{\Pi} = - \hat{S}_n$.

\item The interaction-based readout $U_2$ conserves the parity with respect to the measurement basis, $[U_2,\hat{\Pi}]=0$.

\end{enumerate}

\noindent Thus, it is always possible to saturate the QCRB conditioned on suitable choices of 
measurement basis, readout and phase generator. The conditions (1,2) can be used to determine the optimal measurement basis, which is typically easily accessible. Note that even for the trivial case of readout (ie, $U_2 = 1$), so long as the conditions (1,2) hold, the sensitivity saturates the QCRB \cite{FQnote}.
Nevertheless, a non-trivial choice of $U_2$ (ie, $U_2 \neq 1$) often increases the robustness against detection noise. In this spirit, we are allowed to pick the best readout strategy which satisfies the parity conservation requirement (the third condition). 

In the following Section, we consider the TNT Hamiltonian and demonstrate that the conditions for saturating the QCRB are satisfied. 

\begin{figure}[!ht]
\includegraphics[width=0.45\textwidth]{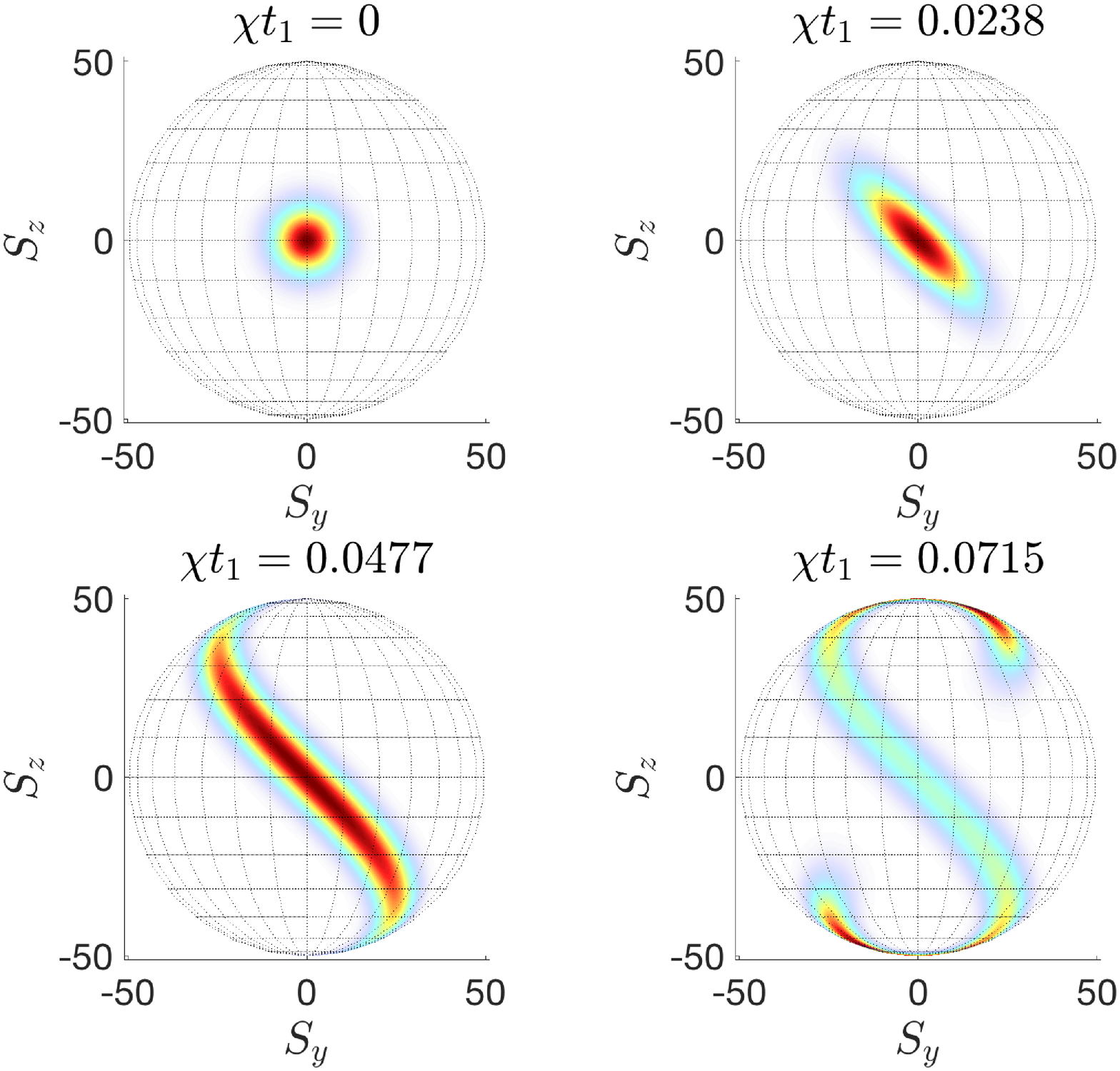}
\includegraphics[width=0.45\textwidth]{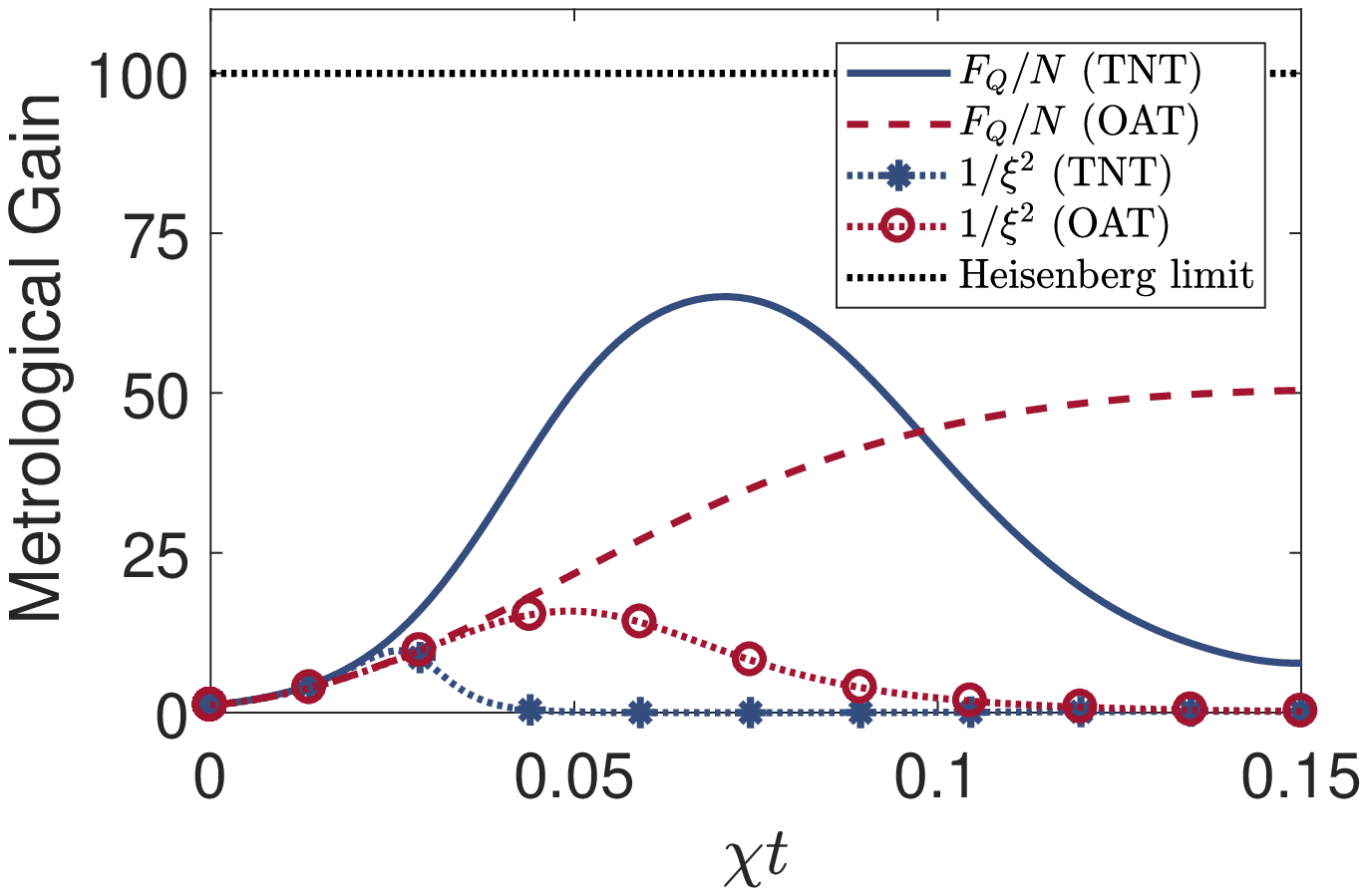} 
\caption{(Color online) Upper panel: Husimi $Q$ function for the state $|\psi\rangle = \exp ({-it}H_{\mathrm{TNT}}/\hbar)|\psi_0\rangle$ under TNT with $\Lambda=2$ and $N = 100$, where $|\psi_0\rangle$ is chosen as the eigenstate of $\hat{S}_x$ with minimum eigenvalue. From left to right: $\chi t =0,\ 0.0238,\ 0.0477$, and $0.0715$. The final frame is the time at which the QFI is maximum. Bloch-sphere plots display the $SU(2)$ Husimi $Q$-function $Q/Q_{\mathrm{max}}$, which is defined 
$Q(\theta, \varphi) = \langle\theta,\varphi|\rho|\theta,\varphi\rangle$, where $|\theta,\varphi\rangle = e^{i \varphi \hat{S}_z}e^{i \theta \hat{S}_y} |S_z = N/2\rangle$ represents the spin coherent 
state along $\theta$ and $\varphi$ directions corresponding to rotating the maximal $\hat{S}_z$ eigenstate around azimuthal  and polar angles $\{\theta, \varphi\}$ \cite{Arecchi:1972}. Lower panel: Metrological gain $F_Q/N$ as a function of $\chi t$ for TNT (blue solid) and OAT (red dashed). We have also included the gain based on spin-squeezing parameter for TNT (blue solid-pentagram) and OAT (red solid-circled). The 
Heisenberg limit ($\Delta \phi = 1/N$) is indicated by the black dotted line.  The initial state is chosen to be the minimal eigenstate of $S_x$.}
     \label{fig:bloch_sphere1}
   \end{figure}

\section{Twist-and-turn interferometry}
We consider the TNT Hamiltonian  \cite{Micheli:2003, Strobel:2014, Muessel:2015} 
\begin{eqnarray}\label{H}
H_{\mathrm{TNT}}=\hbar \chi \hat{S}_z^2-\hbar J \hat{S}_x 
\end{eqnarray}

\noindent such that $U_1 = \exp ( -i t_1 H_{\mathrm{TNT}}/\hbar)$, where the adjustable parameter $t_1$ is the state preparation time. Here, the collective spin operators obey the usual SU(2) commutation relations: $[\hat{S}_i,\hat{S}_j] = \epsilon_{ijk} \hat{S}_k$ where $\epsilon_{ijk}$ is the Levi-Civita symbol. Moreover, $\chi$ and $J$ denote the magnitude of the spin-spin interaction and the rate of rotation about the $\hat{S}_x$ axis, respectively. One can alter the macroscopic properties of the system by adjusting the parameter $\Lambda = N \chi/J$ \cite{smerzi:1997, Milburn:1997}, where $N$ is the total number of particles. Specifically, it has been shown that $\Lambda = 2$, corresponding to maximal criticality of the unstable fixed point in mean-field approximation, provides the maximum rate of entanglement generation \cite{Micheli:2003, Sorelli:2015}. In the limit of $\chi\gg J$ the TNT Hamiltonian reduces to the well-known one-axis-twisting form $H_{\mathrm{OAT}} = \hbar\chi \hat{S}_z^2$.

The TNT Hamiltonian can accelerate the rate of entanglement generation compared to the OAT Hamiltonian for the same $\chi t_1$ \cite{Muessel:2015}. This is the direct consequence of interplay between twist ($\propto \hat{S}_z^2$) and turn ($\propto \hat{S}_x$) terms of the Hamiltonian. The generation of spin-squeezing under TNT dynamics has been investigated 
theoretically \cite{Micheli:2003,Jin:2007,Strobel:2014,Muessel:2015,Sorelli:2015} and realized experimentally in atomic Bose-Einstein condensates \cite{Muessel:2015,Strobel:2014} and cold atomic ensembles \cite{Chaudhury:2007}. 

Figure (\ref{fig:bloch_sphere1}) shows the time evolution of the $SU(2)$ Husimi $Q$-function \cite{Arecchi:1972, Agarwal:1998}, and the QFI under TNT dynamics. The QFI is defined as $F_Q =4\Delta S_n^2$, where $\hat{S}_n$ is the collective spin operator pointing in the direction that maximises the QFI \cite{Hyluss:2010}. However, the QFI is silent on what measurement choice saturates the QCRB. If the $\phi$ estimate is obtained from the mean spin component $\langle \hat{S}_{n^\prime}\rangle$, the error propagation formula gives $\Delta \phi^2 = \Delta S_{n'}^2/\partial_\phi \langle \hat{S}_{n'}\rangle = \xi^2/N$, where $\xi^2 =  N \Delta S_{n'}^2/\langle \hat{S}_\parallel\rangle^2$ is the spin-squeezing parameter normal to mean spin direction $\hat{S}_\parallel $\cite{Wineland:1992}. For small values of $\chi t_1$, this method of estimation is very close to the QCRB, but breaks down when the state becomes non-Gaussian. Non-Gaussian states with $F_Q>N$ are called entangled non-Gaussian states (ENGS), and an analysis of their CFI reveals their metrological gain can exceed that of spin-squeezed states \cite{Strobel:2014, Haine:2015b, Macri:2016,Pezze:2009,Davis:2016}.

\begin{figure}[t]
\centering {\includegraphics[width=\columnwidth]{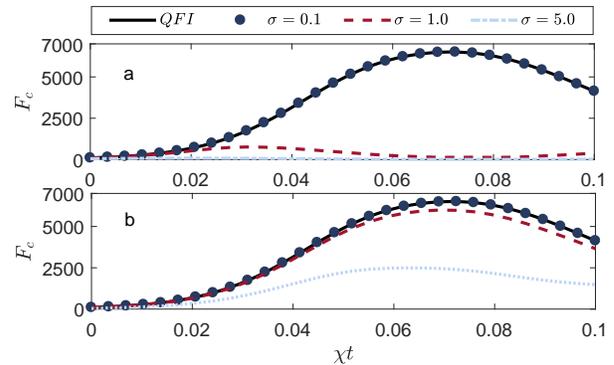}} 
\caption{(Color online) Time evolution of the QFI (solid black) and CFI for
various detection noises $\sigma=0.1$ (dotted blue), $1.0$ (dashed red) and $5.0$ (dotted-dashed light turquoise) for the final state $U_2 U_\phi U_1|\psi_0\rangle$, for (a): $U_2 = 1$, and (b): $U_2 = U_1^\dag$. In all cases the initial state is chosen to be the minimal eigenstate of $S_x$ and
we have fixed $\phi=10^{-4}$. }
\label{fig:FCnoise1}
\end{figure}

We now consider the requirements for performing a measurement that saturates the QCRB with a state generated via TNT dynamics. Assuming the initial state is an eigenstate of $\hat{S}_x$, then $H_{\mathrm{TNT}}$ couples only to $\hat{S}_x$ eigenstates of the same parity. That is, $U_1$ conserves parity with respect to the $\hat{S}_x$ eigenbasis $\{ |m\rangle_x\}$ \cite{Ribeiro:2007,Caneva:2012}. Subsequently, we align the interferometer along the optimal $F_Q$ direction such that $U_{\phi}=\exp(i\hat{S}_n\phi)$. It has been shown that this operator is in the $y-z$ plane, ie, $\hat{S}_n=\alpha \hat{S}_y+\beta \hat{S}_z$ normal to the mean spin direction $\hat{S}_x$ \cite{Sorelli:2015}. Since the generator of the interferometer $\hat{S}_n$ flips the parity (ie, $\hat{\Pi} \hat{S}_n \hat{\Pi} = - \hat{S}_n$, for $\hat{\Pi} = \sum_m |m\rangle_x\langle m|_x$), we see that a measurement that projects into the $\hat{S}_x$ eigenbasis will saturate the QCRB. This is applicable to all readouts that conserve parity with respect to $\hat{S}_x$ (see Appendix \ref{parity} for details).

\begin{figure*}[t]
\centering 
\includegraphics[width =\textwidth]{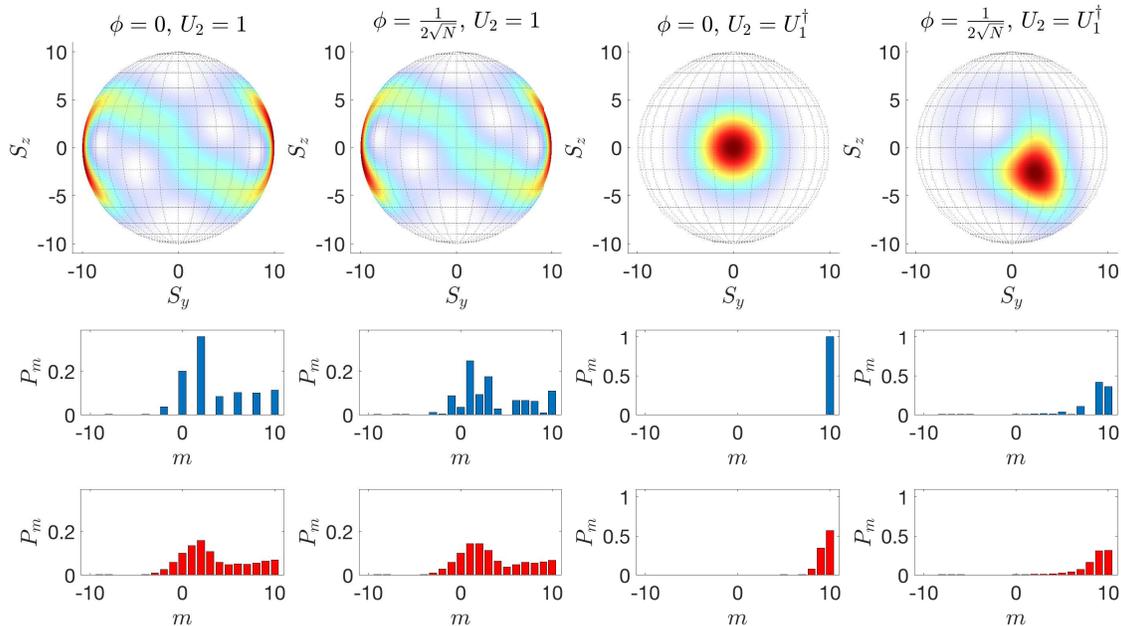}
\caption{(Color online) Top row: Husimi $Q$ function for the final state for (from left to right): $\phi=0$, $U_2=1$; $\phi=\frac{1}{2\sqrt{N}}$, $U_2 = 1$; $\phi = 0$, $U_2= U_1^\dag$; $\phi = \frac{1}{2\sqrt{N}}$, $U_2 = U_1^\dag$. Middle row (blue histograms): The corresponding probabilities $P_m$ for measurements in the $S_x$ eigenbasis for these same states. Bottom row (red histograms): Probabilities in the $S_x$ eigenbasis convolved with detection noise with $\sigma = 1$. Parameters: $N=20$, 
$\Lambda=2$ and $\chi t = 0.275$, which maximises $F_Q$. Note that for reasons of visual clarity, we have rotated the state around the $S_x$ axis such that $S_n = S_y$.}
\label{fig:spinprobs}
\end{figure*}

In realistic situations however, there is detection noise which limits the estimation sensitivity. Both spin-squeezed states and ENGS are more sensitive to detection noise than un-entangled states of the same size, and ENGS typically demand detection noise at the single-atom level \cite{Pezze:2016_review, Davis:2017, Nolan:2017b}, which restricts them to small numbers as the requisite counting efficiency rapidly becomes impossible. For this reason, detection noise is a key limitation in current experiments \cite{Pezze:2016_review}.

The effect of detection noise is to introduce additional, classical noise to the measurement process. For example, a detector that measures the spin projection along $\hat{S}_x$ should always read $S_x = N/2$ for a maximal $\hat{S}_x$ eigenstate. However, if there is noise introduced, there is a finite chance that the detector will read $S_x = m \neq N/2$. To model detection noise we follow \cite{Pezze:2013, Pezze:2016_review, Nolan:2017b} and take the convolution of the probability distribution with a Gaussian distribution with detection noise $\sigma$,
\begin{equation}
\tilde{P}_m(\phi) = \sum_{m^\prime} C_{m^\prime}(\sigma) \exp \left[ -(m-m^\prime)^2/2\sigma^2 \right] P_{m^\prime}(\phi) .
\end{equation}
Imposing $\sum_m \tilde{P}_m(\phi)=1$ one obtains the appropriate normalisation factor,
\begin{equation}
C_{m^\prime}(\sigma) = \left( \sum_{m} \exp\left[-(m-m^\prime)^2/2\sigma^2 \right] \right)^{-1}. 
\end{equation}
We note that this is equivalent to introducing the positive-operator-valued measurement (POVM) \cite{Walls:2008}
\begin{eqnarray}\label{convol} \nonumber
\{\hat{M}_m\}&=& \{|m\rangle\langle m|\} \\
&=&\sum_{m^{\prime}}C_{m^\prime}(\sigma)e^{-(m-m^\prime)^2/2\sigma^2}
|m^{\prime}\rangle\langle m^{\prime}|
\end{eqnarray}
such that the probability distribution of making measurements on the noisey POVMs is given by $\tilde{P}_m(\phi)=\mathrm{Tr}(\hat{M}_{m}U_2U_{\phi}\rho U_{\phi}^{\dagger}U_{2}^{\dagger})$. Note that in the limit of negligible
detection noise ($\sigma\rightarrow 0$) the POVMs approach the orthogonal basis (Section \ref{protocol}).
%
%
In Figure (\ref{fig:FCnoise1}a) we show the CFI calculated from this convolved probability distribution $\tilde{P}_m$ for the state $U_\phi U_1 |\psi_0\rangle$, for different values of $\sigma$. For $\sigma \lesssim 0.1$, our measurement saturates the QCRB. However, for moderate values of $\chi t_1$, a small increase in detection noise significantly degrades the sensitivity. Fortunately, by adding an interaction-based readout such that the final state is  $U_2 U_\phi U_1 |\psi_0\rangle$, with $U_2 = U_1^\dag$, the sensitivity is much more robust to the presence of noise [Figure (\ref{fig:FCnoise1}b)].

We can understand why the interaction-based echo makes the system so much more robust by looking at the $Q$-function and the probability distributions. Figure (\ref{fig:spinprobs}) shows the $Q$-functions and $S_x$ probability distribution for states with $\phi =0$ and a small phase shift $\delta\phi$, for the case with ($U_2 = U_1^\dag$) and without ($U_2 = 1$) an interaction-based readout. In the absence of detection noise, the fidelity between the state with $\phi=0$ and $\delta\phi$ is identical with and without the interaction-based readout. Similarly, the Hellinger distance between the distributions is also identical. The Hellinger distance defines a statistical distance between $P_m(\phi_1)$ and $P_m(\phi_2)$, defined as $d_H^2(\phi_1, \phi_2) = 1-\sum_m \sqrt{P_m(\phi_1)P_m(\phi_2)}$. Adding detection noise, it becomes difficult to distinguish the distributions when $U_2=1$. Compare this to $U_2 = U_1^\dag$, and the two distributions are more easily distinguished. Quantitatively, without the interaction-based readout, the Hellinger distance is $d_H^2 = 0.01$, but when we include the readout it is significantly larger: $d_H^2 = 0.17$ (compared to $d_H^2 = 0.40$ in both cases when no noise is included). However, there is no guarantee that $U_2 = U_1^\dag$ is the best choice of interaction-based readout. In the next section, we examine several possibilities to determine what is the best choice for maximising robustness in TNT squeezing. 

\section{Robustifying entanglement against detection noise}
\noindent We now examine the robustness to detection noise, for a different choices of $U_2$, all satisfying the conditions for optimality. 
In particular, we chose the trivial case of no interaction-based readout ($U_2 = 1$), and the simple time-reversal read-out ($U_2 = U_1^\dag(t_2)$). The latter choice includes asymmetric echo where $t_2 \neq t_1$, which implies $U_2 \neq U_1^\dag$. In this case, we have increased the interaction time for the read-out compared to the state preparation step. Moreover, We also include $U_2 = U_1$, which may be applicable in the case when it is not easy to reverse the sign of the interaction constant $\chi$, such as when one is working with bright-solitons \cite{Haine:2017}, or enhanced nonlinear interactions due to state-dependent potentials \cite{Riedel:2010}. 

\begin{figure}[!ht]
\includegraphics[width=0.45\textwidth]{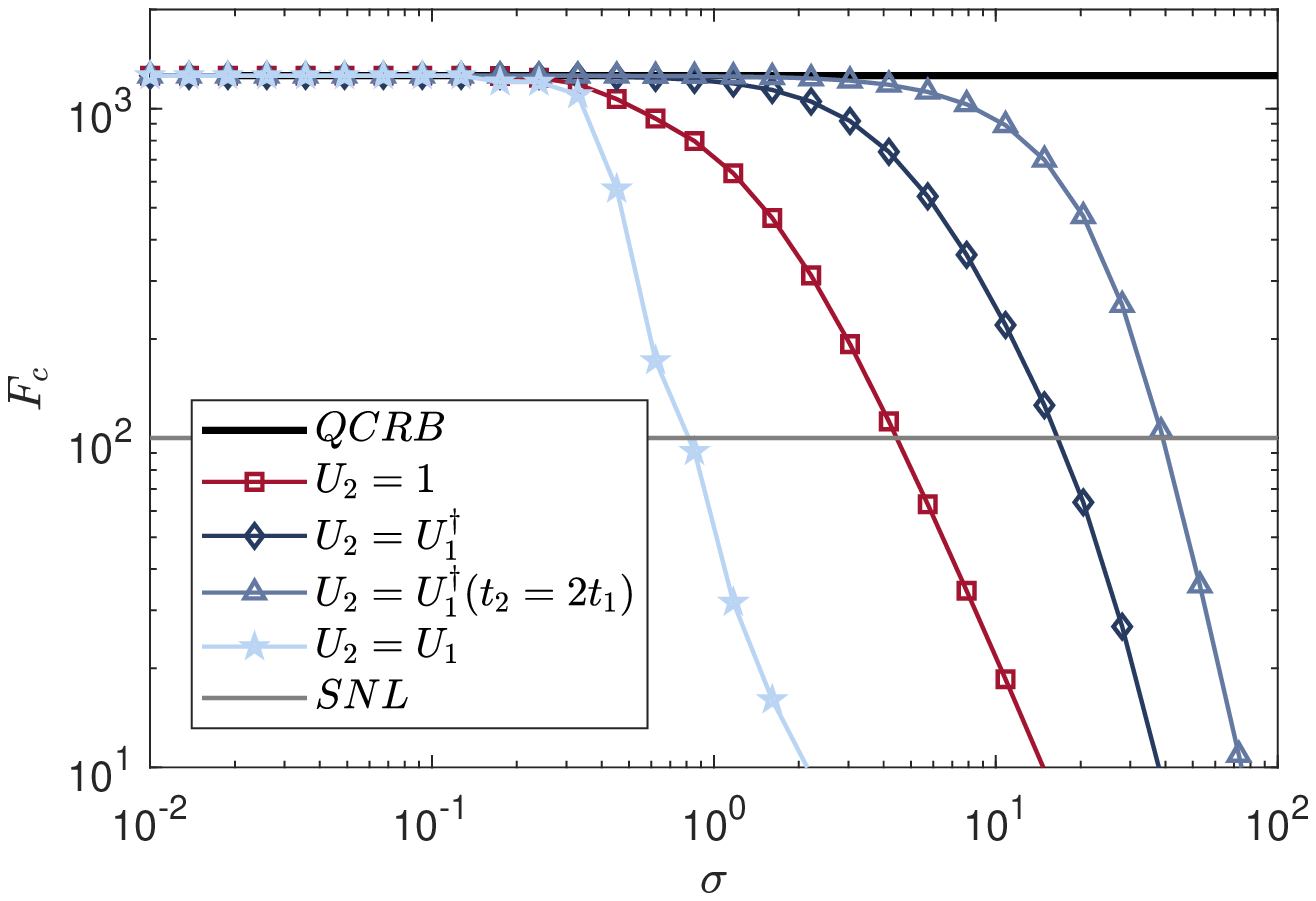}
\includegraphics[width=0.45\textwidth]{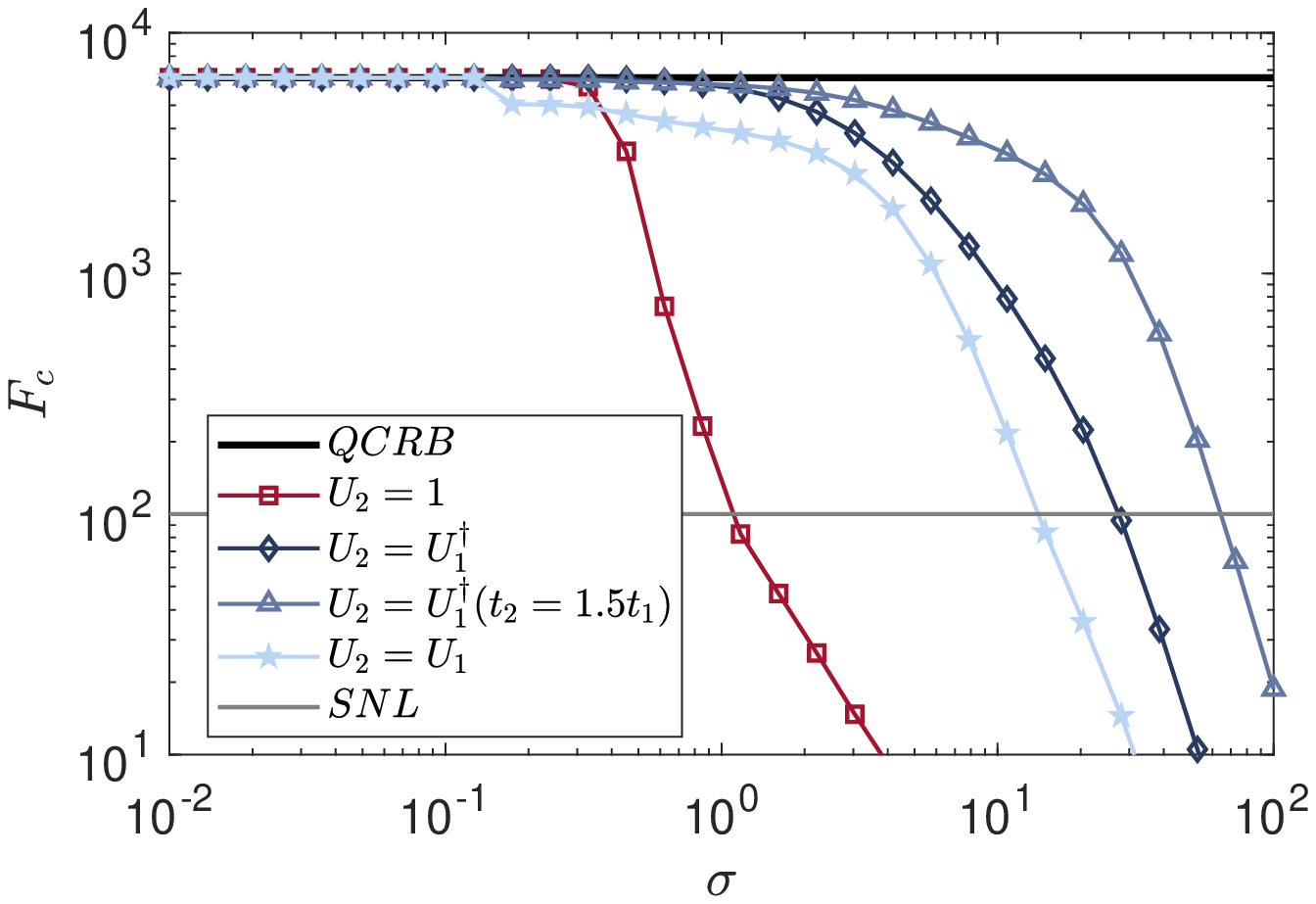}
\caption{(Color online) Decay of classical Fisher information for a many-body entangled
state produced by TNT Hamiltonian in presence of detection noise $\sigma$ for $N=100$, $\Lambda=2$
and $\chi t_1=0.027$ (upper panel) and $\chi t_1=0.072$ (lower panel). We have considered the trivial protocol $U_2=1$ (red squares), an echo $U_2=U_1^\dagger$ (dark blue diamonds), an asymmetric echo $U_2\neq U_1^\dagger$ (light blue triangles) and a pseudo-echo $U_2= U_1$ (light turquoise pentagrams). We have also considered QCRB and SNL as thick black and thin grey lines respectively. Here, $\phi=10^{-4}$. 
}
     \label{fig:Fc_vs_sigma}
   \end{figure}

Figure \ref{fig:Fc_vs_sigma} shows $F_c$ as a function of detection noise for different choices of $U_2$ when $\chi t_1=0.027$ (when the metrological gain using the spin squeezing parameter is maximum) and $0.072$ (when the QFI is maximum). For the case of weak entanglement ($\chi t_1 = 0.027$), we see that the trivial case ($U_2 = 1$) is degraded to below shot-noise levels for noise of approximately $\sigma \approx 4.40$. However, by adding the `echo' unitary ($U_2 = U_1^\dag$), sub-shot-noise sensitivities are retained up to $\sigma \approx 16.45$. Interestingly the choice $U_2 \neq U_1^\dag$ provides greater robustness again, with sub-shot-noise sensitivities with up to $\sigma = 39.17$, which is significantly greater than $\sqrt{N}$. For this value of $\chi t_1$, $U_2 = U_1$ provides no advantage compared to the trivial case. However, for the case of maximum QFI ($\chi t = 0.072$), this choice does provide some additional robustness compared to the trivial case. The asymmetric echo provides the greatest robustness of the schemes considered, and therefore may be useful when there is unavoidably large detection noise. To obtain $F_c$, we have numerically optimised the measurement basis in the planes normal to $\hat{S}_x$ and $\hat{S}_n$ \cite{note2}.  Based on our numerical calculations, it seems that by increasing the detection noise,
the optimal measurement basis moves into the normal plane to $S_x$. 

\begin{figure}[t]
\centering {\includegraphics[width=8.0 cm]{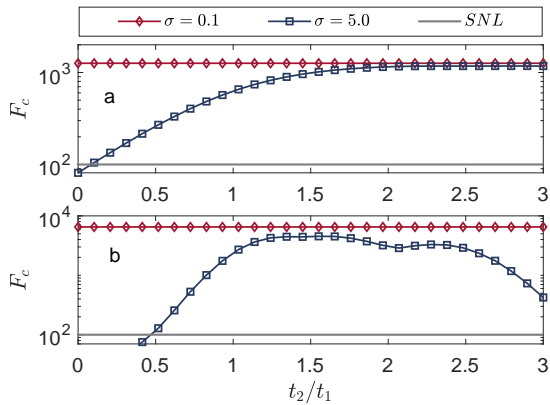}}
\caption{(Color online) The Fisher information produced by TNT versus time duration of echo $t_2/t_1$ for $U_2=U_1^\dagger(t_2)$ in presence of detection noise $\sigma=0.1$ (red diamonds) and $5.0$ (blue squares) when (a): spin squeezing is maximum $t_1=0.027$ and (b): when the QFI is maximum $t_1=0.072$. For both cases, when there is negligible detection noise, the CFI corresponds to QCRB as expected. The SNL is the grey line.}
\label{fig:figure5}
\end{figure}

We now more closely examine the effect of an asymmetric echo on the TNT scheme. In Figure \ref{fig:figure5}, we have changed the time duration of the echo $t_2/t_1$ for fixed detection noises $\sigma=0.1,5.0$ when the
TNT Hamiltonian produces the maximum value of spin squeezing ($t_1=0.027$) or maximum value of quantum Fisher information ($t_1=0.072$) respectively. For small values of detection noise $\sigma=0.1$, $U_2$ does not affect the classical Fisher information.
In this case, as mentioned before, $F_c$ saturates the QCRB. In contrast, for more noise ($\sigma=5.0$) we see that in the Gaussian regime ($t_1=0.027$), by increasing the duration of the echo ($t_2/t_1$) the sensitivity becomes more robust against detection noise and eventually approaches the QCRB value for large enough $t_2$. Compare this to the non-Gaussian regime ($t_1=0.072$), where there is a clear optimum readout time of roughly $t_2/t_1 \approx 1.5$. Recently, similar behaviour has been reported in two-axis-counter-twisted interaction-based readout scheme \cite{Anders:2018}.

\begin{figure}[t]
\centering {\includegraphics[width=8.0 cm]{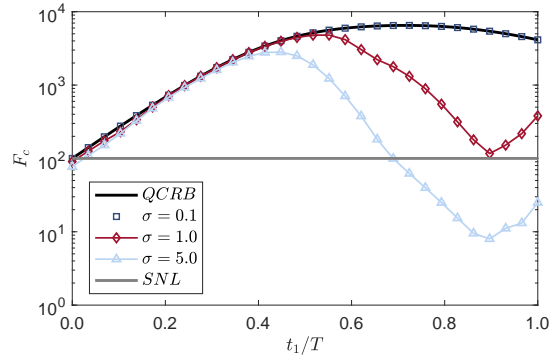}}
\caption{(Color online) Classical Fisher information versus entangling duration $t_1$, with fixed total experimental time
$\chi T=\chi(t_1+t_2)=0.1$. The TNT entanglement and re-entangling operation [$U_2= U_1^\dag(t_2)$] are applied for durations $t_1$ and $t_2 = T-t_1$ respectively, such that $|\psi_\phi \rangle = U_1^\dagger(t_2) U_\phi U_1(t_1) |\psi_0 \rangle$). Here, $N=100$ and $\Lambda=2$ and we have considered $\sigma=0.1$ (dark blue squares), $1.0$ (red diamonds) and $\sigma=5.0$ (light turquoise triangles). The QCRB and SNL are given with solid thick black and thin grey lines. 
}
\label{fig:figure6}
\end{figure}

In a realistic spin-squeezing set-up, the duration of the experiment may be limited by the
particle losses as well as dephasing \cite{Li:2008, Li:2009} and multi-mode dynamics \cite{Haine:2009, Haine:2014}. Therefore, it is important to consider the optimal portion of time for entangling ($U_1$) and re-entangling ($U_2$) sequences, as there is a trade-off between the desire to maximise the QFI via the state preparation operation $U_1(t_1)$, and robustifying against detection noise via the interaction-based readout operation $U_2(t_2)$, while keeping the total time fixed $T = t_1 + t_2$.  In Figure \ref{fig:figure6} we have given the 
optimized Fisher information versus entangling unitary time for total experimental duration $T=0.1$ with $N=100$ and $\Lambda=2$ and time reversal echo case $U_2=U_1^{\dagger}$. When the detection noise is small ($\sigma=0.1$), there is little benefit in increasing $t_2$ and the optimum strategy is to choose $U_2=1$. However this is certainly not true for large detection noise, as devoting more time to the entangling operation compromises the interaction-based readout. Therefore, to maximise the CFI there is a balance to be found between twisting the initial state for longer (increasing $t_1$) and robustifying the state (increasing $t_2$). 
For larger values of detection noise ($\sigma=1.0$, $5.0$) we see that the optimal strategy is close to an echo ($t_1 \approx T/2=0.050$). The reason is that when we devote less than half of the time to the TNT entangling operation,
there is always a possibility to perform an echo with $t_2/t_1\geq 1$. Thus, the decrease in $F_c$ as a result of
reducing the first TNT unitary is compensated for by increased robustness provided by the second unitary. For instance, Figure \ref{fig:figure5}a gives the typical behaviour of the asymmetric echo within this region, which approaches to the QCRB as $t_2/t_1$ increases beyond $1$. On the other hand, by increasing the portion of the entangling duration more than $T/2$, the echo time ratio $t_2/t_1$ decreases below unity, and is far more susceptible to detection noise. 

\section{Conclusion}\label{conclu}
Many-body entangled states are a crucial resources for quantum-enhanced metrology. Current experiments are working to devise schemes that are able to rapidly manufacture these states in large atomic ensembles, but detection noise is currently a key limitation \cite{Pezze:2016_review}. The TNT interaction is capable of generating entanglement faster than the conventional one-axis-twisting interaction. TNT is capable of rapidly producing both spin-squeezed states and ENGS. In this work we explore the use of interaction-based readout protocols to provide rapid quantum-enhanced metrology based on twist-and-turn dynamics, which is also robust to detection noise and optimally utilises the state's entanglement. This is done with a spin-resolving measurement in the optimal basis, which we provide criteria for determining.  In this regard, we have confirmed the usefulness of standard symmetric echo protocols in boosting the measurement performance against noise. However, 
our results imply that for weakly entangled initial states, using an asymmetric echo provides better robustness than
the symmetric case. Finally, we have considered TNT echo protocols in realistic situation where there is a limitation on the total time allowed for both the initial interaction and the readout. We have shown that the best outcomes require balance between entangling and re-entangling 
time durations. When detection noise is small, any readout is sub-optimal. In presence of considerable noise, the optimal strategy is close to an echo, but the precise time trade-off depends on the magnitude of detection noise present in the system.

\section{Acknowledgements} The authors acknowledge useful discussions with Augusto Smerzi, Luca Pezze, and Manuel Gessner. Numerical simulations were performed on the University of Queensland School of Mathematics and Physics computing cluster ``Dogmatix,'' with thanks to I.~Mortimer for computing support. S.P.N. acknowledges support provided by an Australian Postgraduate Award. S.A.H. has received funding from the European Union's Horizon 2020 research and innovation programme under the Marie Sklodowska-Curie Grant Agreement No.~704672.

\appendix
\section{The conditions of QCRB saturation} \label{sec:appendixa}
\noindent  Here, we present our conditions for saturation of QCRB (Section 2) 
when applying spin-resolving measurement \cite{Nolan:2017b}.

To begin, lets ignore the re-entangling readout operator, and consider only $U_2=1$. After the state preparation and phase encoding steps, the probability of estimating the small, unknown phase $\phi$ is given by 
$P_m(\phi)=\langle m|U_{\phi}\rho U_{\phi}^\dagger|m\rangle$ with $U_{\phi}=e^{(-iS_n\phi)}$. Our phase estimation is limited by the CFI. For small values of $\phi$ the CFI can be approximated as the leading term in the expansion of the Hellinger statistical distance in the space of probability distributions\cite{Strobel:2014}
\begin{eqnarray}\nonumber
d_H^2(0,\phi)&=&1-\sum_{m}\sqrt{P_m(0)P_m(\phi)}\\ \label{Hll}
&=&F_c(0)\phi^2/8+\mathcal{O}(\phi^3)
\end{eqnarray}

\noindent For small $\phi$, Taylor expanding the probability amplitude
gives

\begin{eqnarray}
\nonumber
P_m(\phi)&=&P_m(0)+ \frac{\partial P_m(\phi)}{\partial\phi}\bigg|_{\phi=0}\phi
+ \frac{\partial^2_{\phi}P_m(\phi)}{\partial\phi^2}\bigg|_{\phi=0}\frac{\phi^2}{2}\\ \label{tay}
&+&\mathcal{O}(\phi^3) .
\end{eqnarray}

\noindent We have

\begin{eqnarray}
\nonumber
P_m(\phi)&=&\langle m|U_{\phi}^\dagger\rho U_{\phi}|m\rangle,\\ \nonumber
\frac{\partial P_m(\phi)}{\partial\phi}&=&i\langle m|\hat{S}_nU_{\phi}\rho U_{\phi}^{\dagger}|m\rangle +c.c.\\ \nonumber
\frac{\partial^2 P_m(\phi)}{\partial\phi^2}&=&\langle m|\hat{S}_nU_{\phi}\rho U_{\phi}^{\dagger}\hat{S}_n|m\rangle \\ 
 &-&\langle m|\hat{S}_n^2U_{\phi}\rho U_{\phi}^{\dagger}|m\rangle + c.c.
\end{eqnarray}

\noindent which leads to
\begin{eqnarray}
\nonumber
P_m(0)&=&\langle m|\rho|m\rangle, \\ \nonumber
\frac{\partial P_m(\phi)}{\partial\phi}\bigg|_{\phi=0}&=&i\langle m|\hat{S}_n\rho|m\rangle +c.c.\\ \nonumber
\frac{\partial^2 P_m(\phi)}{\partial\phi^2}\bigg|_{\phi=0}&=&\langle m|\hat{S}_n\rho \hat{S}_n|m\rangle \\ 
 &-&\langle m|\hat{S}_n^2\rho|m\rangle + c.c.
\end{eqnarray}

\noindent Now, suppose that the density operator is an eigenstate of the parity operator $\hat{\Pi}=\sum_m(-1)^m |m\rangle \langle m|$
such that $\hat{\Pi}\rho=(-1)^p\rho$ for $p=0,1$. Also, let $\hat{\Pi} \hat{S}_n \hat{\Pi}=-\hat{S}_n$ (Section \ref{protocol}). Under these two conditions,
\begin{eqnarray}
\nonumber
\langle m|\rho|m\rangle&=&(-1)^{m+p}\langle m|\rho|m\rangle, \\ \nonumber
\langle m|\rho \hat{S}_n|m\rangle&=&0, \\ \label{avG0}
\langle m|S_n \rho \hat{S}_n|m\rangle &=& (-1)^{m+p+1}\langle m|\hat{S}_n \rho \hat{S}_n|m\rangle,
\end{eqnarray}

\noindent which also yields
\begin{eqnarray}\label{or}
 P_m(0)\langle m|\hat{S}_n \rho \hat{S}_n|m\rangle=0, 
\end{eqnarray}
as $P_m(0) = 0$ if $m+p$ is odd, and $\langle m|\hat{S}_n \rho \hat{S}_n|m\rangle = 0$ if $m+p$ is even.  After using \ref{tay} and \ref{or} in \ref{Hll} followed by a binomial expansion of square root for small $\phi$, we obtain

\begin{eqnarray}
d_H^2(\phi)=\sum_m \langle m|\hat{S}_n^2\rho|m\rangle/2
\end{eqnarray}

\noindent up to third order in $\phi$. Finally, as $\langle \hat{S}_n \rangle=0$ [\ref{avG0}],
up to third order in $\phi$ the Fisher information is  
\begin{eqnarray}
\nonumber
F_c(0)&=&4\sum_m\langle m|\hat{S}_n^2\rho|m\rangle\\ \nonumber
&=&4\Delta {\hat{S}_n^2}\\
&=&F_Q
\end{eqnarray}

\noindent where, $\Delta{S_n^2}$ is the variance of generator. The last equality
appears since $F_c\leq F_Q \leq \Delta S_n^2$. Thus, the first two parity conditions 
(section \ref{protocol}) ensure the saturation of the QCRB. There is no need to use the re-entangling step to obtain this result. However, In order to robustify the scheme against noise,
we include the readout operator $U_2$ after phase encoding which gives, $P_m(\phi)=\langle m|U_2^{\dagger} U_{\phi}^{\dagger} \rho U_{\phi} U_2|m\rangle$. This probability distribution has equivalent CFI when compared to the distribution $P^\prime_m =\langle m|U_{\phi}^{\dagger} \rho U_{\phi}|m\rangle$
if $[U_2,\hat{\Pi}]=0$, i.e. the readout operator has the same parity as 
measurement basis $|m\rangle$ (condition 3 given in section \ref{protocol}).

\section{Fulfilment of parity conditions for TNT scenario}\label{parity}
The TNT Hamiltonian conserves parity, $[H_{\mathrm{TNT}},\hat{\Pi}_x]=0$ \cite{Ribeiro:2007,Caneva:2012}. Consequently by taking the initial state as $\hat{S}_x$ eigenstate, the first parity condition of section \ref{protocol}
is satisfied.  Moreover, the parity symmetry  leads to $\hat{\Pi}_x(\hat{S}_x,\hat{S}_y,\hat{S}_z)\hat{\Pi}_x=(\hat{S}_x,-\hat{S}_y,-\hat{S}_z)$. Considering the fact that optimal QFI is in the $y-z$ plane \cite{Strobel:2014,Sorelli:2015} $\hat{S}_n=\alpha \hat{S}_y+\beta \hat{S}_z$, one obtains 
$\hat{\Pi}_x \hat{S}_n \hat{\Pi}_x=-\hat{S}_n$ which demonstrates the second condition. To demonstrate the third condition, we again note that the re-entangling unitary conserves parity in the $\hat{S}_x$ eigenbasis. Of course, asymmetry in the entangling and re-entangling operations does not affect this result.

\bibliography{../echo_bib}

\begin{thebibliography}{49}%
\makeatletter
\providecommand \@ifxundefined [1]{%
 \@ifx{#1\undefined}
}%
\providecommand \@ifnum [1]{%
 \ifnum #1\expandafter \@firstoftwo
 \else \expandafter \@secondoftwo
 \fi
}%
\providecommand \@ifx [1]{%
 \ifx #1\expandafter \@firstoftwo
 \else \expandafter \@secondoftwo
 \fi
}%
\providecommand \natexlab [1]{#1}%
\providecommand \enquote  [1]{``#1''}%
\providecommand \bibnamefont  [1]{#1}%
\providecommand \bibfnamefont [1]{#1}%
\providecommand \citenamefont [1]{#1}%
\providecommand \href@noop [0]{\@secondoftwo}%
\providecommand \href [0]{\begingroup \@sanitize@url \@href}%
\providecommand \@href[1]{\@@startlink{#1}\@@href}%
\providecommand \@@href[1]{\endgroup#1\@@endlink}%
\providecommand \@sanitize@url [0]{\catcode `\\12\catcode `\$12\catcode
  `\&12\catcode `\#12\catcode `\^12\catcode `\_12\catcode `\%12\relax}%
\providecommand \@@startlink[1]{}%
\providecommand \@@endlink[0]{}%
\providecommand \url  [0]{\begingroup\@sanitize@url \@url }%
\providecommand \@url [1]{\endgroup\@href {#1}{\urlprefix }}%
\providecommand \urlprefix  [0]{URL }%
\providecommand \Eprint [0]{\href }%
\providecommand \doibase [0]{http://dx.doi.org/}%
\providecommand \selectlanguage [0]{\@gobble}%
\providecommand \bibinfo  [0]{\@secondoftwo}%
\providecommand \bibfield  [0]{\@secondoftwo}%
\providecommand \translation [1]{[#1]}%
\providecommand \BibitemOpen [0]{}%
\providecommand \bibitemStop [0]{}%
\providecommand \bibitemNoStop [0]{.\EOS\space}%
\providecommand \EOS [0]{\spacefactor3000\relax}%
\providecommand \BibitemShut  [1]{\csname bibitem#1\endcsname}%
\let\auto@bib@innerbib\@empty
\bibitem [{\citenamefont {Pezze}\ \emph {et~al.}(2016)\citenamefont {Pezze},
  \citenamefont {Smerzi}, \citenamefont {Oberthaler}, \citenamefont {Schmied},\
  and\ \citenamefont {Treutlein}}]{Pezze:2016_review}%
  \BibitemOpen
  \bibfield  {author} {\bibinfo {author} {\bibfnamefont {L.}~\bibnamefont
  {Pezze}}, \bibinfo {author} {\bibfnamefont {A.}~\bibnamefont {Smerzi}},
  \bibinfo {author} {\bibfnamefont {M.~K.}\ \bibnamefont {Oberthaler}},
  \bibinfo {author} {\bibfnamefont {R.}~\bibnamefont {Schmied}}, \ and\
  \bibinfo {author} {\bibfnamefont {P.}~\bibnamefont {Treutlein}},\ }\bibfield
  {title} {\enquote {\bibinfo {title} {Quantum metrology with nonclassical
  states of atomic ensembles},}\ }\href@noop {} {\bibfield  {journal} {\bibinfo
   {journal} {arXiv:1609.01609}\ } (\bibinfo {year} {2016})}\BibitemShut
  {NoStop}%
\bibitem [{\citenamefont {Giovannetti}\ \emph {et~al.}(2006)\citenamefont
  {Giovannetti}, \citenamefont {Lloyd},\ and\ \citenamefont
  {Maccone}}]{Giovannetti:2006}%
  \BibitemOpen
  \bibfield  {author} {\bibinfo {author} {\bibfnamefont {Vittorio}\
  \bibnamefont {Giovannetti}}, \bibinfo {author} {\bibfnamefont {Seth}\
  \bibnamefont {Lloyd}}, \ and\ \bibinfo {author} {\bibfnamefont {Lorenzo}\
  \bibnamefont {Maccone}},\ }\bibfield  {title} {\enquote {\bibinfo {title}
  {Quantum metrology},}\ }\href {\doibase 10.1103/PhysRevLett.96.010401}
  {\bibfield  {journal} {\bibinfo  {journal} {Phys. Rev. Lett.}\ }\textbf
  {\bibinfo {volume} {96}},\ \bibinfo {pages} {010401} (\bibinfo {year}
  {2006})}\BibitemShut {NoStop}%
\bibitem [{\citenamefont {Pezz\'e}\ and\ \citenamefont
  {Smerzi}(2009)}]{Pezze:2009}%
  \BibitemOpen
  \bibfield  {author} {\bibinfo {author} {\bibfnamefont {Luca}\ \bibnamefont
  {Pezz\'e}}\ and\ \bibinfo {author} {\bibfnamefont {Augusto}\ \bibnamefont
  {Smerzi}},\ }\bibfield  {title} {\enquote {\bibinfo {title} {Entanglement,
  nonlinear dynamics, and the {Heisenberg} limit},}\ }\href {\doibase
  10.1103/PhysRevLett.102.100401} {\bibfield  {journal} {\bibinfo  {journal}
  {Phys. Rev. Lett.}\ }\textbf {\bibinfo {volume} {102}},\ \bibinfo {pages}
  {100401} (\bibinfo {year} {2009})}\BibitemShut {NoStop}%
\bibitem [{\citenamefont {Kitagawa}\ and\ \citenamefont
  {Ueda}(1993)}]{Kitagawa:1993}%
  \BibitemOpen
  \bibfield  {author} {\bibinfo {author} {\bibfnamefont {Masahiro}\
  \bibnamefont {Kitagawa}}\ and\ \bibinfo {author} {\bibfnamefont {Masahito}\
  \bibnamefont {Ueda}},\ }\bibfield  {title} {\enquote {\bibinfo {title}
  {Squeezed spin states},}\ }\href {\doibase 10.1103/PhysRevA.47.5138}
  {\bibfield  {journal} {\bibinfo  {journal} {Phys. Rev. A}\ }\textbf {\bibinfo
  {volume} {47}},\ \bibinfo {pages} {5138--5143} (\bibinfo {year}
  {1993})}\BibitemShut {NoStop}%
\bibitem [{\citenamefont {M\o{}lmer}\ and\ \citenamefont
  {S\o{}rensen}(1999)}]{Molmer:1999}%
  \BibitemOpen
  \bibfield  {author} {\bibinfo {author} {\bibfnamefont {Klaus}\ \bibnamefont
  {M\o{}lmer}}\ and\ \bibinfo {author} {\bibfnamefont {Anders}\ \bibnamefont
  {S\o{}rensen}},\ }\bibfield  {title} {\enquote {\bibinfo {title}
  {Multiparticle entanglement of hot trapped ions},}\ }\href {\doibase
  10.1103/PhysRevLett.82.1835} {\bibfield  {journal} {\bibinfo  {journal}
  {Phys. Rev. Lett.}\ }\textbf {\bibinfo {volume} {82}},\ \bibinfo {pages}
  {1835--1838} (\bibinfo {year} {1999})}\BibitemShut {NoStop}%
\bibitem [{\citenamefont {Esteve}\ \emph {et~al.}(2008)\citenamefont {Esteve},
  \citenamefont {Gross}, \citenamefont {A.}, \citenamefont {Giovanazzi},\ and\
  \citenamefont {Oberthaler}}]{Esteve:2008}%
  \BibitemOpen
  \bibfield  {author} {\bibinfo {author} {\bibfnamefont {J.}~\bibnamefont
  {Esteve}}, \bibinfo {author} {\bibfnamefont {C.}~\bibnamefont {Gross}},
  \bibinfo {author} {\bibfnamefont {Weller.}\ \bibnamefont {A.}}, \bibinfo
  {author} {\bibfnamefont {S.}~\bibnamefont {Giovanazzi}}, \ and\ \bibinfo
  {author} {\bibfnamefont {M.~K.}\ \bibnamefont {Oberthaler}},\ }\bibfield
  {title} {\enquote {\bibinfo {title} {Squeezing and entanglement in a
  {Bose-Einstein} condensate},}\ }\href {\doibase 10.1038/nature07332}
  {\bibfield  {journal} {\bibinfo  {journal} {Nature}\ }\textbf {\bibinfo
  {volume} {455}},\ \bibinfo {pages} {1216} (\bibinfo {year}
  {2008})}\BibitemShut {NoStop}%
\bibitem [{\citenamefont {Leroux}\ \emph {et~al.}(2010)\citenamefont {Leroux},
  \citenamefont {Schleier-Smith},\ and\ \citenamefont
  {Vuleti\ifmmode~\acute{c}\else \'{c}\fi{}}}]{Leroux:2010}%
  \BibitemOpen
  \bibfield  {author} {\bibinfo {author} {\bibfnamefont {Ian~D.}\ \bibnamefont
  {Leroux}}, \bibinfo {author} {\bibfnamefont {Monika~H.}\ \bibnamefont
  {Schleier-Smith}}, \ and\ \bibinfo {author} {\bibfnamefont {Vladan}\
  \bibnamefont {Vuleti\ifmmode~\acute{c}\else \'{c}\fi{}}},\ }\bibfield
  {title} {\enquote {\bibinfo {title} {Implementation of cavity squeezing of a
  collective atomic spin},}\ }\href {\doibase 10.1103/PhysRevLett.104.073602}
  {\bibfield  {journal} {\bibinfo  {journal} {Phys. Rev. Lett.}\ }\textbf
  {\bibinfo {volume} {104}},\ \bibinfo {pages} {073602} (\bibinfo {year}
  {2010})}\BibitemShut {NoStop}%
\bibitem [{\citenamefont {Gross}\ \emph {et~al.}(2010)\citenamefont {Gross},
  \citenamefont {Zibold}, \citenamefont {Nicklas}, \citenamefont {Est{\`e}ve},\
  and\ \citenamefont {Oberthaler}}]{Gross:2010}%
  \BibitemOpen
  \bibfield  {author} {\bibinfo {author} {\bibfnamefont {C.}~\bibnamefont
  {Gross}}, \bibinfo {author} {\bibfnamefont {T.}~\bibnamefont {Zibold}},
  \bibinfo {author} {\bibfnamefont {E.}~\bibnamefont {Nicklas}}, \bibinfo
  {author} {\bibfnamefont {J.}~\bibnamefont {Est{\`e}ve}}, \ and\ \bibinfo
  {author} {\bibfnamefont {M.~K.}\ \bibnamefont {Oberthaler}},\ }\bibfield
  {title} {\enquote {\bibinfo {title} {Nonlinear atom interferometer surpasses
  classical precision limit},}\ }\href {http://dx.doi.org/10.1038/nature08919}
  {\bibfield  {journal} {\bibinfo  {journal} {Nature}\ }\textbf {\bibinfo
  {volume} {464}},\ \bibinfo {pages} {1165--1169} (\bibinfo {year}
  {2010})}\BibitemShut {NoStop}%
\bibitem [{\citenamefont {Riedel}\ \emph {et~al.}(2010)\citenamefont {Riedel},
  \citenamefont {B\"ohi}, \citenamefont {Li}, \citenamefont {H\"ansch},
  \citenamefont {Sinatra},\ and\ \citenamefont {Treutlein}}]{Riedel:2010}%
  \BibitemOpen
  \bibfield  {author} {\bibinfo {author} {\bibfnamefont {Max~F.}\ \bibnamefont
  {Riedel}}, \bibinfo {author} {\bibfnamefont {Pascal}\ \bibnamefont {B\"ohi}},
  \bibinfo {author} {\bibfnamefont {Yun}\ \bibnamefont {Li}}, \bibinfo {author}
  {\bibfnamefont {Theodor~W.}\ \bibnamefont {H\"ansch}}, \bibinfo {author}
  {\bibfnamefont {Alice}\ \bibnamefont {Sinatra}}, \ and\ \bibinfo {author}
  {\bibfnamefont {Philipp}\ \bibnamefont {Treutlein}},\ }\bibfield  {title}
  {\enquote {\bibinfo {title} {Atom-chip-based generation of entanglement for
  quantum metrology},}\ }\href {http://dx.doi.org/10.1038/nature08988}
  {\bibfield  {journal} {\bibinfo  {journal} {Nature}\ }\textbf {\bibinfo
  {volume} {464}},\ \bibinfo {pages} {1170--1173} (\bibinfo {year}
  {2010})}\BibitemShut {NoStop}%
\bibitem [{\citenamefont {Muessel}\ \emph {et~al.}(2014)\citenamefont
  {Muessel}, \citenamefont {Strobel}, \citenamefont {Linnemann}, \citenamefont
  {Hume},\ and\ \citenamefont {Oberthaler}}]{Muessel:2014}%
  \BibitemOpen
  \bibfield  {author} {\bibinfo {author} {\bibfnamefont {W.}~\bibnamefont
  {Muessel}}, \bibinfo {author} {\bibfnamefont {H.}~\bibnamefont {Strobel}},
  \bibinfo {author} {\bibfnamefont {D.}~\bibnamefont {Linnemann}}, \bibinfo
  {author} {\bibfnamefont {D.~B.}\ \bibnamefont {Hume}}, \ and\ \bibinfo
  {author} {\bibfnamefont {M.~K.}\ \bibnamefont {Oberthaler}},\ }\bibfield
  {title} {\enquote {\bibinfo {title} {Scalable spin squeezing for
  quantum-enhanced magnetometry with {Bose-Einstein} condensates},}\ }\href
  {\doibase 10.1103/PhysRevLett.113.103004} {\bibfield  {journal} {\bibinfo
  {journal} {Phys. Rev. Lett.}\ }\textbf {\bibinfo {volume} {113}},\ \bibinfo
  {pages} {103004} (\bibinfo {year} {2014})}\BibitemShut {NoStop}%
\bibitem [{\citenamefont {Li}\ \emph {et~al.}(2008)\citenamefont {Li},
  \citenamefont {Castin},\ and\ \citenamefont {Sinatra}}]{Li:2008}%
  \BibitemOpen
  \bibfield  {author} {\bibinfo {author} {\bibfnamefont {Yun}\ \bibnamefont
  {Li}}, \bibinfo {author} {\bibfnamefont {Y.}~\bibnamefont {Castin}}, \ and\
  \bibinfo {author} {\bibfnamefont {A.}~\bibnamefont {Sinatra}},\ }\bibfield
  {title} {\enquote {\bibinfo {title} {Optimum spin squeezing in
  {Bose-Einstein} condensates with particle losses},}\ }\href {\doibase
  10.1103/PhysRevLett.100.210401} {\bibfield  {journal} {\bibinfo  {journal}
  {Phys. Rev. Lett.}\ }\textbf {\bibinfo {volume} {100}},\ \bibinfo {pages}
  {210401} (\bibinfo {year} {2008})}\BibitemShut {NoStop}%
\bibitem [{\citenamefont {Li}\ \emph {et~al.}(2009)\citenamefont {Li},
  \citenamefont {Treutlein}, \citenamefont {Reichel},\ and\ \citenamefont
  {Sinatra}}]{Li:2009}%
  \BibitemOpen
  \bibfield  {author} {\bibinfo {author} {\bibfnamefont {Yun}\ \bibnamefont
  {Li}}, \bibinfo {author} {\bibfnamefont {P.}~\bibnamefont {Treutlein}},
  \bibinfo {author} {\bibfnamefont {J.}~\bibnamefont {Reichel}}, \ and\
  \bibinfo {author} {\bibfnamefont {A.}~\bibnamefont {Sinatra}},\ }\bibfield
  {title} {\enquote {\bibinfo {title} {Spin squeezing in a bimodal condensate:
  spatial dynamics and particle losses},}\ }\href {\doibase
  10.1140/epjb/e2008-00472-6} {\bibfield  {journal} {\bibinfo  {journal} {The
  European Physical Journal B}\ }\textbf {\bibinfo {volume} {68}},\ \bibinfo
  {pages} {365--381} (\bibinfo {year} {2009})}\BibitemShut {NoStop}%
\bibitem [{\citenamefont {Haine}\ and\ \citenamefont
  {Johnsson}(2009)}]{Haine:2009}%
  \BibitemOpen
  \bibfield  {author} {\bibinfo {author} {\bibfnamefont {Simon~A.}\
  \bibnamefont {Haine}}\ and\ \bibinfo {author} {\bibfnamefont {Mattias~T.}\
  \bibnamefont {Johnsson}},\ }\bibfield  {title} {\enquote {\bibinfo {title}
  {Dynamic scheme for generating number squeezing in {B}ose-{E}instein
  condensates through nonlinear interactions},}\ }\href {\doibase
  10.1103/PhysRevA.80.023611} {\bibfield  {journal} {\bibinfo  {journal} {Phys.
  Rev. A}\ }\textbf {\bibinfo {volume} {80}},\ \bibinfo {pages} {023611}
  (\bibinfo {year} {2009})}\BibitemShut {NoStop}%
\bibitem [{\citenamefont {Haine}\ \emph {et~al.}(2014)\citenamefont {Haine},
  \citenamefont {Lau}, \citenamefont {Anderson},\ and\ \citenamefont
  {Johnsson}}]{Haine:2014}%
  \BibitemOpen
  \bibfield  {author} {\bibinfo {author} {\bibfnamefont {S.~A.}\ \bibnamefont
  {Haine}}, \bibinfo {author} {\bibfnamefont {J.}~\bibnamefont {Lau}}, \bibinfo
  {author} {\bibfnamefont {R.~P.}\ \bibnamefont {Anderson}}, \ and\ \bibinfo
  {author} {\bibfnamefont {M.~T.}\ \bibnamefont {Johnsson}},\ }\bibfield
  {title} {\enquote {\bibinfo {title} {Self-induced spatial dynamics to enhance
  spin squeezing via one-axis twisting in a two-component {Bose-Einstein}
  condensate},}\ }\href {\doibase 10.1103/PhysRevA.90.023613} {\bibfield
  {journal} {\bibinfo  {journal} {Phys. Rev. A}\ }\textbf {\bibinfo {volume}
  {90}},\ \bibinfo {pages} {023613} (\bibinfo {year} {2014})}\BibitemShut
  {NoStop}%
\bibitem [{\citenamefont {Micheli}\ \emph {et~al.}(2003)\citenamefont
  {Micheli}, \citenamefont {Jaksch}, \citenamefont {Cirac},\ and\ \citenamefont
  {Zoller}}]{Micheli:2003}%
  \BibitemOpen
  \bibfield  {author} {\bibinfo {author} {\bibfnamefont {A.}~\bibnamefont
  {Micheli}}, \bibinfo {author} {\bibfnamefont {D.}~\bibnamefont {Jaksch}},
  \bibinfo {author} {\bibfnamefont {J.~I.}\ \bibnamefont {Cirac}}, \ and\
  \bibinfo {author} {\bibfnamefont {P.}~\bibnamefont {Zoller}},\ }\bibfield
  {title} {\enquote {\bibinfo {title} {Many-particle entanglement in
  two-component {Bose-Einstein} condensates},}\ }\href {\doibase
  10.1103/PhysRevA.67.013607} {\bibfield  {journal} {\bibinfo  {journal} {Phys.
  Rev. A}\ }\textbf {\bibinfo {volume} {67}},\ \bibinfo {pages} {013607}
  (\bibinfo {year} {2003})}\BibitemShut {NoStop}%
\bibitem [{\citenamefont {Strobel}\ \emph {et~al.}(2014)\citenamefont
  {Strobel}, \citenamefont {Muessel}, \citenamefont {Linnemann}, \citenamefont
  {Zibold}, \citenamefont {Hume}, \citenamefont {Pezz{\`e}}, \citenamefont
  {Smerzi},\ and\ \citenamefont {Oberthaler}}]{Strobel:2014}%
  \BibitemOpen
  \bibfield  {author} {\bibinfo {author} {\bibfnamefont {H.}~\bibnamefont
  {Strobel}}, \bibinfo {author} {\bibfnamefont {W.}~\bibnamefont {Muessel}},
  \bibinfo {author} {\bibfnamefont {D.}~\bibnamefont {Linnemann}}, \bibinfo
  {author} {\bibfnamefont {T.}~\bibnamefont {Zibold}}, \bibinfo {author}
  {\bibfnamefont {D.~B.}\ \bibnamefont {Hume}}, \bibinfo {author}
  {\bibfnamefont {L.}~\bibnamefont {Pezz{\`e}}}, \bibinfo {author}
  {\bibfnamefont {A.}~\bibnamefont {Smerzi}}, \ and\ \bibinfo {author}
  {\bibfnamefont {M.~K.}\ \bibnamefont {Oberthaler}},\ }\bibfield  {title}
  {\enquote {\bibinfo {title} {Fisher information and entanglement of
  non-{G}aussian spin states},}\ }\href {\doibase 10.1126/science.1250147}
  {\bibfield  {journal} {\bibinfo  {journal} {Science}\ }\textbf {\bibinfo
  {volume} {345}},\ \bibinfo {pages} {424--427} (\bibinfo {year}
  {2014})}\BibitemShut {NoStop}%
\bibitem [{\citenamefont {Muessel}\ \emph {et~al.}(2015)\citenamefont
  {Muessel}, \citenamefont {Strobel}, \citenamefont {Linnemann}, \citenamefont
  {Zibold}, \citenamefont {Juli\'a-D\'{\i}az},\ and\ \citenamefont
  {Oberthaler}}]{Muessel:2015}%
  \BibitemOpen
  \bibfield  {author} {\bibinfo {author} {\bibfnamefont {W.}~\bibnamefont
  {Muessel}}, \bibinfo {author} {\bibfnamefont {H.}~\bibnamefont {Strobel}},
  \bibinfo {author} {\bibfnamefont {D.}~\bibnamefont {Linnemann}}, \bibinfo
  {author} {\bibfnamefont {T.}~\bibnamefont {Zibold}}, \bibinfo {author}
  {\bibfnamefont {B.}~\bibnamefont {Juli\'a-D\'{\i}az}}, \ and\ \bibinfo
  {author} {\bibfnamefont {M.~K.}\ \bibnamefont {Oberthaler}},\ }\bibfield
  {title} {\enquote {\bibinfo {title} {Twist-and-turn spin squeezing in
  {Bose-Einstein} condensates},}\ }\href {\doibase 10.1103/PhysRevA.92.023603}
  {\bibfield  {journal} {\bibinfo  {journal} {Phys. Rev. A}\ }\textbf {\bibinfo
  {volume} {92}},\ \bibinfo {pages} {023603} (\bibinfo {year}
  {2015})}\BibitemShut {NoStop}%
\bibitem [{\citenamefont {Sorelli}\ \emph {et~al.}(2015)\citenamefont
  {Sorelli}, \citenamefont {Pezze},\ and\ \citenamefont
  {Smerzi}}]{Sorelli:2015}%
  \BibitemOpen
  \bibfield  {author} {\bibinfo {author} {\bibfnamefont {G}~\bibnamefont
  {Sorelli}}, \bibinfo {author} {\bibfnamefont {L}~\bibnamefont {Pezze}}, \
  and\ \bibinfo {author} {\bibfnamefont {A}~\bibnamefont {Smerzi}},\ }\bibfield
   {title} {\enquote {\bibinfo {title} {Fast generation of entanglement in
  bosonic {Josephson} junctions},}\ }\href@noop {} {\bibfield  {journal}
  {\bibinfo  {journal} {arXiv:1511.01835}\ } (\bibinfo {year}
  {2015})}\BibitemShut {NoStop}%
\bibitem [{\citenamefont {Marino}\ \emph {et~al.}(2012)\citenamefont {Marino},
  \citenamefont {Corzo~Trejo},\ and\ \citenamefont {Lett}}]{Marino:2012}%
  \BibitemOpen
  \bibfield  {author} {\bibinfo {author} {\bibfnamefont {A.~M.}\ \bibnamefont
  {Marino}}, \bibinfo {author} {\bibfnamefont {N.~V.}\ \bibnamefont
  {Corzo~Trejo}}, \ and\ \bibinfo {author} {\bibfnamefont {P.~D.}\ \bibnamefont
  {Lett}},\ }\bibfield  {title} {\enquote {\bibinfo {title} {Effect of losses
  on the performance of an \uppercase{SU}(1,1) interferometer},}\ }\href
  {\doibase 10.1103/PhysRevA.86.023844} {\bibfield  {journal} {\bibinfo
  {journal} {Phys. Rev. A}\ }\textbf {\bibinfo {volume} {86}},\ \bibinfo
  {pages} {023844} (\bibinfo {year} {2012})}\BibitemShut {NoStop}%
\bibitem [{\citenamefont {Hosten}\ \emph {et~al.}(2016)\citenamefont {Hosten},
  \citenamefont {Krishnakumar}, \citenamefont {Engelsen},\ and\ \citenamefont
  {Kasevich}}]{Hosten:2016}%
  \BibitemOpen
  \bibfield  {author} {\bibinfo {author} {\bibfnamefont {O.}~\bibnamefont
  {Hosten}}, \bibinfo {author} {\bibfnamefont {R.}~\bibnamefont
  {Krishnakumar}}, \bibinfo {author} {\bibfnamefont {N.~J.}\ \bibnamefont
  {Engelsen}}, \ and\ \bibinfo {author} {\bibfnamefont {M.~A.}\ \bibnamefont
  {Kasevich}},\ }\bibfield  {title} {\enquote {\bibinfo {title} {Quantum phase
  magnification},}\ }\href {\doibase 10.1126/science.aaf3397} {\bibfield
  {journal} {\bibinfo  {journal} {Science}\ }\textbf {\bibinfo {volume}
  {352}},\ \bibinfo {pages} {1552--1555} (\bibinfo {year} {2016})}\BibitemShut
  {NoStop}%
\bibitem [{\citenamefont {Davis}\ \emph {et~al.}(2016)\citenamefont {Davis},
  \citenamefont {Bentsen},\ and\ \citenamefont {Schleier-Smith}}]{Davis:2016}%
  \BibitemOpen
  \bibfield  {author} {\bibinfo {author} {\bibfnamefont {E.}~\bibnamefont
  {Davis}}, \bibinfo {author} {\bibfnamefont {G.}~\bibnamefont {Bentsen}}, \
  and\ \bibinfo {author} {\bibfnamefont {M.}~\bibnamefont {Schleier-Smith}},\
  }\bibfield  {title} {\enquote {\bibinfo {title} {Approaching the heisenberg
  limit without single-particle detection},}\ }\href {\doibase
  10.1103/PhysRevLett.116.053601} {\bibfield  {journal} {\bibinfo  {journal}
  {Phys. Rev. Lett.}\ }\textbf {\bibinfo {volume} {116}},\ \bibinfo {pages}
  {053601} (\bibinfo {year} {2016})}\BibitemShut {NoStop}%
\bibitem [{\citenamefont {Macr\`{\i}}\ \emph {et~al.}(2016)\citenamefont
  {Macr\`{\i}}, \citenamefont {Smerzi},\ and\ \citenamefont
  {Pezz\`e}}]{Macri:2016}%
  \BibitemOpen
  \bibfield  {author} {\bibinfo {author} {\bibfnamefont {Tommaso}\ \bibnamefont
  {Macr\`{\i}}}, \bibinfo {author} {\bibfnamefont {Augusto}\ \bibnamefont
  {Smerzi}}, \ and\ \bibinfo {author} {\bibfnamefont {Luca}\ \bibnamefont
  {Pezz\`e}},\ }\bibfield  {title} {\enquote {\bibinfo {title} {Loschmidt echo
  for quantum metrology},}\ }\href {\doibase 10.1103/PhysRevA.94.010102}
  {\bibfield  {journal} {\bibinfo  {journal} {Phys. Rev. A}\ }\textbf {\bibinfo
  {volume} {94}},\ \bibinfo {pages} {010102} (\bibinfo {year}
  {2016})}\BibitemShut {NoStop}%
\bibitem [{\citenamefont {Fr\"owis}\ \emph {et~al.}(2016)\citenamefont
  {Fr\"owis}, \citenamefont {Sekatski},\ and\ \citenamefont
  {D\"ur}}]{Frowis:2016}%
  \BibitemOpen
  \bibfield  {author} {\bibinfo {author} {\bibfnamefont {Florian}\ \bibnamefont
  {Fr\"owis}}, \bibinfo {author} {\bibfnamefont {Pavel}\ \bibnamefont
  {Sekatski}}, \ and\ \bibinfo {author} {\bibfnamefont {Wolfgang}\ \bibnamefont
  {D\"ur}},\ }\bibfield  {title} {\enquote {\bibinfo {title} {Detecting large
  quantum {Fisher} information with finite measurement precision},}\ }\href
  {\doibase 10.1103/PhysRevLett.116.090801} {\bibfield  {journal} {\bibinfo
  {journal} {Phys. Rev. Lett.}\ }\textbf {\bibinfo {volume} {116}},\ \bibinfo
  {pages} {090801} (\bibinfo {year} {2016})}\BibitemShut {NoStop}%
\bibitem [{\citenamefont {Szigeti}\ \emph {et~al.}(2017)\citenamefont
  {Szigeti}, \citenamefont {Lewis-Swan},\ and\ \citenamefont
  {Haine}}]{Szigeti:2017}%
  \BibitemOpen
  \bibfield  {author} {\bibinfo {author} {\bibfnamefont {Stuart~S.}\
  \bibnamefont {Szigeti}}, \bibinfo {author} {\bibfnamefont {Robert~J.}\
  \bibnamefont {Lewis-Swan}}, \ and\ \bibinfo {author} {\bibfnamefont
  {Simon~A.}\ \bibnamefont {Haine}},\ }\bibfield  {title} {\enquote {\bibinfo
  {title} {Pumped-up \uppercase{SU}(1,1) interferometry},}\ }\href {\doibase
  10.1103/PhysRevLett.118.150401} {\bibfield  {journal} {\bibinfo  {journal}
  {Phys. Rev. Lett.}\ }\textbf {\bibinfo {volume} {118}},\ \bibinfo {pages}
  {150401} (\bibinfo {year} {2017})}\BibitemShut {NoStop}%
\bibitem [{\citenamefont {Davis}\ \emph {et~al.}(2017)\citenamefont {Davis},
  \citenamefont {Bentsen}, \citenamefont {Li},\ and\ \citenamefont
  {Schleier-Smith}}]{Davis:2017}%
  \BibitemOpen
  \bibfield  {author} {\bibinfo {author} {\bibfnamefont {Emily}\ \bibnamefont
  {Davis}}, \bibinfo {author} {\bibfnamefont {Gregory}\ \bibnamefont
  {Bentsen}}, \bibinfo {author} {\bibfnamefont {Tracy}\ \bibnamefont {Li}}, \
  and\ \bibinfo {author} {\bibfnamefont {Monika}\ \bibnamefont
  {Schleier-Smith}},\ }\bibfield  {title} {\enquote {\bibinfo {title}
  {Advantages of interaction-based readout for quantum sensing},}\ \ }(\bibinfo
  {year} {2017})\ pp.\ \bibinfo {pages} {101180Z--101180Z--10}\BibitemShut
  {NoStop}%
\bibitem [{\citenamefont {Nolan}\ \emph {et~al.}(2017)\citenamefont {Nolan},
  \citenamefont {Szigeti},\ and\ \citenamefont {Haine}}]{Nolan:2017b}%
  \BibitemOpen
  \bibfield  {author} {\bibinfo {author} {\bibfnamefont {Samuel~P.}\
  \bibnamefont {Nolan}}, \bibinfo {author} {\bibfnamefont {Stuart~S.}\
  \bibnamefont {Szigeti}}, \ and\ \bibinfo {author} {\bibfnamefont {Simon~A.}\
  \bibnamefont {Haine}},\ }\bibfield  {title} {\enquote {\bibinfo {title}
  {Optimal and robust quantum metrology using interaction-based readouts},}\
  }\href {\doibase 10.1103/PhysRevLett.119.193601} {\bibfield  {journal}
  {\bibinfo  {journal} {Phys. Rev. Lett.}\ }\textbf {\bibinfo {volume} {119}},\
  \bibinfo {pages} {193601} (\bibinfo {year} {2017})}\BibitemShut {NoStop}%
\bibitem [{\citenamefont {Huang}\ \emph {et~al.}(2018)\citenamefont {Huang},
  \citenamefont {Zhuang},\ and\ \citenamefont {Lee}}]{Huang:2017}%
  \BibitemOpen
  \bibfield  {author} {\bibinfo {author} {\bibfnamefont {Jiahao}\ \bibnamefont
  {Huang}}, \bibinfo {author} {\bibfnamefont {Min}\ \bibnamefont {Zhuang}}, \
  and\ \bibinfo {author} {\bibfnamefont {Chaohong}\ \bibnamefont {Lee}},\
  }\bibfield  {title} {\enquote {\bibinfo {title} {Non-{G}aussian precision
  metrology via driving through quantum phase transitions},}\ }\href {\doibase
  10.1103/PhysRevA.97.032116} {\bibfield  {journal} {\bibinfo  {journal} {Phys.
  Rev. A}\ }\textbf {\bibinfo {volume} {97}},\ \bibinfo {pages} {032116}
  (\bibinfo {year} {2018})}\BibitemShut {NoStop}%
\bibitem [{\citenamefont {Anders}\ \emph {et~al.}(2018)\citenamefont {Anders},
  \citenamefont {Pezz\`e}, \citenamefont {Smerzi},\ and\ \citenamefont
  {Klempt}}]{Anders:2018}%
  \BibitemOpen
  \bibfield  {author} {\bibinfo {author} {\bibfnamefont {Fabian}\ \bibnamefont
  {Anders}}, \bibinfo {author} {\bibfnamefont {Luca}\ \bibnamefont {Pezz\`e}},
  \bibinfo {author} {\bibfnamefont {Augusto}\ \bibnamefont {Smerzi}}, \ and\
  \bibinfo {author} {\bibfnamefont {Carsten}\ \bibnamefont {Klempt}},\
  }\bibfield  {title} {\enquote {\bibinfo {title} {Phase magnification by
  two-axis countertwisting for detection-noise robust interferometry},}\ }\href
  {\doibase 10.1103/PhysRevA.97.043813} {\bibfield  {journal} {\bibinfo
  {journal} {Phys. Rev. A}\ }\textbf {\bibinfo {volume} {97}},\ \bibinfo
  {pages} {043813} (\bibinfo {year} {2018})}\BibitemShut {NoStop}%
\bibitem [{\citenamefont {Fang}\ \emph {et~al.}(2017)\citenamefont {Fang},
  \citenamefont {Sarkar},\ and\ \citenamefont {Shahriar}}]{Fang:2017}%
  \BibitemOpen
  \bibfield  {author} {\bibinfo {author} {\bibfnamefont {R}~\bibnamefont
  {Fang}}, \bibinfo {author} {\bibfnamefont {R}~\bibnamefont {Sarkar}}, \ and\
  \bibinfo {author} {\bibfnamefont {S.~M.}\ \bibnamefont {Shahriar}},\
  }\bibfield  {title} {\enquote {\bibinfo {title} {Enhancing sensitivity of an
  atom interferometer to the {Heisenberg} limit using increased quantum
  noise},}\ }\href@noop {} {\bibfield  {journal} {\bibinfo  {journal}
  {arXiv:1707.08260}\ } (\bibinfo {year} {2017})}\BibitemShut {NoStop}%
\bibitem [{\citenamefont {Hayes}\ \emph {et~al.}(2018)\citenamefont {Hayes},
  \citenamefont {Dooley}, \citenamefont {Munro}, \citenamefont {Nemoto},\ and\
  \citenamefont {Dunningham}}]{Hayes:2018}%
  \BibitemOpen
  \bibfield  {author} {\bibinfo {author} {\bibfnamefont {A.~J.}\ \bibnamefont
  {Hayes}}, \bibinfo {author} {\bibfnamefont {S.}~\bibnamefont {Dooley}},
  \bibinfo {author} {\bibfnamefont {W.~J.}\ \bibnamefont {Munro}}, \bibinfo
  {author} {\bibfnamefont {K.}~\bibnamefont {Nemoto}}, \ and\ \bibinfo {author}
  {\bibfnamefont {J.}~\bibnamefont {Dunningham}},\ }\bibfield  {title}
  {\enquote {\bibinfo {title} {Making the most of time in quantum metrology:
  concurrent state preparation and sensing},}\ }\href@noop {} {\bibfield
  {journal} {\bibinfo  {journal} {arXiv:1801.03452}\ } (\bibinfo {year}
  {2018})}\BibitemShut {NoStop}%
\bibitem [{\citenamefont {Braunstein}\ and\ \citenamefont
  {Caves}(1994)}]{Braunstein:1994}%
  \BibitemOpen
  \bibfield  {author} {\bibinfo {author} {\bibfnamefont {Samuel~L.}\
  \bibnamefont {Braunstein}}\ and\ \bibinfo {author} {\bibfnamefont
  {Carlton~M.}\ \bibnamefont {Caves}},\ }\bibfield  {title} {\enquote {\bibinfo
  {title} {Statistical distance and the geometry of quantum states},}\ }\href
  {\doibase 10.1103/PhysRevLett.72.3439} {\bibfield  {journal} {\bibinfo
  {journal} {Phys. Rev. Lett.}\ }\textbf {\bibinfo {volume} {72}},\ \bibinfo
  {pages} {3439--3443} (\bibinfo {year} {1994})}\BibitemShut {NoStop}%
\bibitem [{\citenamefont {T\'oth}\ and\ \citenamefont
  {Apellaniz}(2014)}]{Toth:2014}%
  \BibitemOpen
  \bibfield  {author} {\bibinfo {author} {\bibfnamefont {G\'eza}\ \bibnamefont
  {T\'oth}}\ and\ \bibinfo {author} {\bibfnamefont {Iagoba}\ \bibnamefont
  {Apellaniz}},\ }\bibfield  {title} {\enquote {\bibinfo {title} {Quantum
  metrology from a quantum information science perspective},}\ }\href
  {http://stacks.iop.org/1751-8121/47/i=42/a=424006} {\bibfield  {journal}
  {\bibinfo  {journal} {Journal of Physics A: Mathematical and Theoretical}\
  }\textbf {\bibinfo {volume} {47}},\ \bibinfo {pages} {424006} (\bibinfo
  {year} {2014})}\BibitemShut {NoStop}%
\bibitem [{\citenamefont {Demkowicz-Dobrza{\'n}ski}\ \emph
  {et~al.}(2015)\citenamefont {Demkowicz-Dobrza{\'n}ski}, \citenamefont
  {Jarzyna},\ and\ \citenamefont
  {Ko{\l}ody{\'n}ski}}]{Demkowicz-Dobrzanski:2014}%
  \BibitemOpen
  \bibfield  {author} {\bibinfo {author} {\bibfnamefont {Rafa{\l}}\
  \bibnamefont {Demkowicz-Dobrza{\'n}ski}}, \bibinfo {author} {\bibfnamefont
  {M.}~\bibnamefont {Jarzyna}}, \ and\ \bibinfo {author} {\bibfnamefont
  {J.}~\bibnamefont {Ko{\l}ody{\'n}ski}},\ }\bibfield  {title} {\enquote
  {\bibinfo {title} {Quantum limits in optical interferometry},}\ }\href@noop
  {} {\bibfield  {journal} {\bibinfo  {journal} {Progress in Optics}\ }\textbf
  {\bibinfo {volume} {345}} (\bibinfo {year} {2015})}\BibitemShut {NoStop}%
\bibitem [{FQn()}]{FQnote}%
  \BibitemOpen
  \href@noop {} {}\bibinfo {howpublished} {Also, note that using $U_2$ does not
  change the QFI. This can be easily verified when using the equation $F_Q= 4 [
  \langle \partial_\phi \psi| \partial_\phi \psi\rangle - |\langle
  \psi|\partial_\phi\psi\rangle |^2]$ for $|\psi\rangle=U_2U_\phi
  U_1|\psi_0\rangle$ where $\partial_\phi=\partial/\partial\phi$.}\BibitemShut
  {Stop}%
\bibitem [{\citenamefont {Arecchi}\ \emph {et~al.}(1972)\citenamefont
  {Arecchi}, \citenamefont {Courtens}, \citenamefont {Gilmore},\ and\
  \citenamefont {Thomas}}]{Arecchi:1972}%
  \BibitemOpen
  \bibfield  {author} {\bibinfo {author} {\bibfnamefont {F.~T.}\ \bibnamefont
  {Arecchi}}, \bibinfo {author} {\bibfnamefont {Eric}\ \bibnamefont
  {Courtens}}, \bibinfo {author} {\bibfnamefont {Robert}\ \bibnamefont
  {Gilmore}}, \ and\ \bibinfo {author} {\bibfnamefont {Harry}\ \bibnamefont
  {Thomas}},\ }\bibfield  {title} {\enquote {\bibinfo {title} {Atomic coherent
  states in quantum optics},}\ }\href {\doibase 10.1103/PhysRevA.6.2211}
  {\bibfield  {journal} {\bibinfo  {journal} {Phys. Rev. A}\ }\textbf {\bibinfo
  {volume} {6}},\ \bibinfo {pages} {2211--2237} (\bibinfo {year}
  {1972})}\BibitemShut {NoStop}%
\bibitem [{\citenamefont {Smerzi}\ \emph {et~al.}(1997)\citenamefont {Smerzi},
  \citenamefont {Fantoni}, \citenamefont {Giovanazzi},\ and\ \citenamefont
  {Shenoy}}]{smerzi:1997}%
  \BibitemOpen
  \bibfield  {author} {\bibinfo {author} {\bibfnamefont {A.}~\bibnamefont
  {Smerzi}}, \bibinfo {author} {\bibfnamefont {S.}~\bibnamefont {Fantoni}},
  \bibinfo {author} {\bibfnamefont {S.}~\bibnamefont {Giovanazzi}}, \ and\
  \bibinfo {author} {\bibfnamefont {S.~R.}\ \bibnamefont {Shenoy}},\ }\bibfield
   {title} {\enquote {\bibinfo {title} {Quantum coherent atomic tunneling
  between two trapped {Bose-Einstein} condensates},}\ }\href {\doibase
  10.1103/PhysRevLett.79.4950} {\bibfield  {journal} {\bibinfo  {journal}
  {Phys. Rev. Lett.}\ }\textbf {\bibinfo {volume} {79}},\ \bibinfo {pages}
  {4950--4953} (\bibinfo {year} {1997})}\BibitemShut {NoStop}%
\bibitem [{\citenamefont {Milburn}\ \emph {et~al.}(1997)\citenamefont
  {Milburn}, \citenamefont {Corney}, \citenamefont {Wright},\ and\
  \citenamefont {Walls}}]{Milburn:1997}%
  \BibitemOpen
  \bibfield  {author} {\bibinfo {author} {\bibfnamefont {G.~J.}\ \bibnamefont
  {Milburn}}, \bibinfo {author} {\bibfnamefont {J.}~\bibnamefont {Corney}},
  \bibinfo {author} {\bibfnamefont {E.~M.}\ \bibnamefont {Wright}}, \ and\
  \bibinfo {author} {\bibfnamefont {D.~F.}\ \bibnamefont {Walls}},\ }\bibfield
  {title} {\enquote {\bibinfo {title} {Quantum dynamics of an atomic
  {Bose-Einstein} condensate in a double-well potential},}\ }\href {\doibase
  10.1103/PhysRevA.55.4318} {\bibfield  {journal} {\bibinfo  {journal} {Phys.
  Rev. A}\ }\textbf {\bibinfo {volume} {55}},\ \bibinfo {pages} {4318--4324}
  (\bibinfo {year} {1997})}\BibitemShut {NoStop}%
\bibitem [{\citenamefont {Jin}\ and\ \citenamefont {Kim}(2007)}]{Jin:2007}%
  \BibitemOpen
  \bibfield  {author} {\bibinfo {author} {\bibfnamefont {Guang-Ri}\
  \bibnamefont {Jin}}\ and\ \bibinfo {author} {\bibfnamefont {Sang~Wook}\
  \bibnamefont {Kim}},\ }\bibfield  {title} {\enquote {\bibinfo {title} {Spin
  squeezing and maximal-squeezing time},}\ }\href {\doibase
  10.1103/PhysRevA.76.043621} {\bibfield  {journal} {\bibinfo  {journal} {Phys.
  Rev. A}\ }\textbf {\bibinfo {volume} {76}},\ \bibinfo {pages} {043621}
  (\bibinfo {year} {2007})}\BibitemShut {NoStop}%
\bibitem [{\citenamefont {Chaudhury}\ \emph {et~al.}(2007)\citenamefont
  {Chaudhury}, \citenamefont {Merkel}, \citenamefont {Herr}, \citenamefont
  {Silberfarb}, \citenamefont {Deutsch},\ and\ \citenamefont
  {Jessen}}]{Chaudhury:2007}%
  \BibitemOpen
  \bibfield  {author} {\bibinfo {author} {\bibfnamefont {Souma}\ \bibnamefont
  {Chaudhury}}, \bibinfo {author} {\bibfnamefont {Seth}\ \bibnamefont
  {Merkel}}, \bibinfo {author} {\bibfnamefont {Tobias}\ \bibnamefont {Herr}},
  \bibinfo {author} {\bibfnamefont {Andrew}\ \bibnamefont {Silberfarb}},
  \bibinfo {author} {\bibfnamefont {Ivan~H.}\ \bibnamefont {Deutsch}}, \ and\
  \bibinfo {author} {\bibfnamefont {Poul~S.}\ \bibnamefont {Jessen}},\
  }\bibfield  {title} {\enquote {\bibinfo {title} {Quantum control of the
  hyperfine spin of a {Cs} atom ensemble},}\ }\href@noop {} {\bibfield
  {journal} {\bibinfo  {journal} {Phys. Rev. Lett.}\ }\textbf {\bibinfo
  {volume} {99}},\ \bibinfo {pages} {163002} (\bibinfo {year}
  {2007})}\BibitemShut {NoStop}%
\bibitem [{\citenamefont {Agarwal}(1998)}]{Agarwal:1998}%
  \BibitemOpen
  \bibfield  {author} {\bibinfo {author} {\bibfnamefont {G.~S.}\ \bibnamefont
  {Agarwal}},\ }\bibfield  {title} {\enquote {\bibinfo {title} {State
  reconstruction for a collection of two-level systems},}\ }\href {\doibase
  10.1103/PhysRevA.57.671} {\bibfield  {journal} {\bibinfo  {journal} {Phys.
  Rev. A}\ }\textbf {\bibinfo {volume} {57}},\ \bibinfo {pages} {671--673}
  (\bibinfo {year} {1998})}\BibitemShut {NoStop}%
\bibitem [{\citenamefont {Hyllus}\ \emph {et~al.}(2010)\citenamefont {Hyllus},
  \citenamefont {G\"uhne},\ and\ \citenamefont {Smerzi}}]{Hyluss:2010}%
  \BibitemOpen
  \bibfield  {author} {\bibinfo {author} {\bibfnamefont {Philipp}\ \bibnamefont
  {Hyllus}}, \bibinfo {author} {\bibfnamefont {Otfried}\ \bibnamefont
  {G\"uhne}}, \ and\ \bibinfo {author} {\bibfnamefont {Augusto}\ \bibnamefont
  {Smerzi}},\ }\bibfield  {title} {\enquote {\bibinfo {title} {Not all pure
  entangled states are useful for sub-shot-noise interferometry},}\ }\href
  {\doibase 10.1103/PhysRevA.82.012337} {\bibfield  {journal} {\bibinfo
  {journal} {Phys. Rev. A}\ }\textbf {\bibinfo {volume} {82}},\ \bibinfo
  {pages} {012337} (\bibinfo {year} {2010})}\BibitemShut {NoStop}%
\bibitem [{\citenamefont {Wineland}\ \emph {et~al.}(1992)\citenamefont
  {Wineland}, \citenamefont {Bollinger}, \citenamefont {Itano}, \citenamefont
  {Moore},\ and\ \citenamefont {Heinzen}}]{Wineland:1992}%
  \BibitemOpen
  \bibfield  {author} {\bibinfo {author} {\bibfnamefont {D.~J.}\ \bibnamefont
  {Wineland}}, \bibinfo {author} {\bibfnamefont {J.~J.}\ \bibnamefont
  {Bollinger}}, \bibinfo {author} {\bibfnamefont {W.~M.}\ \bibnamefont
  {Itano}}, \bibinfo {author} {\bibfnamefont {F.~L.}\ \bibnamefont {Moore}}, \
  and\ \bibinfo {author} {\bibfnamefont {D.~J.}\ \bibnamefont {Heinzen}},\
  }\bibfield  {title} {\enquote {\bibinfo {title} {Spin squeezing and reduced
  quantum noise in spectroscopy},}\ }\href {\doibase 10.1103/PhysRevA.46.R6797}
  {\bibfield  {journal} {\bibinfo  {journal} {Phys. Rev. A}\ }\textbf {\bibinfo
  {volume} {46}},\ \bibinfo {pages} {R6797--R6800} (\bibinfo {year}
  {1992})}\BibitemShut {NoStop}%
\bibitem [{\citenamefont {Haine}\ and\ \citenamefont
  {Szigeti}(2015)}]{Haine:2015b}%
  \BibitemOpen
  \bibfield  {author} {\bibinfo {author} {\bibfnamefont {Simon~A.}\
  \bibnamefont {Haine}}\ and\ \bibinfo {author} {\bibfnamefont {Stuart~S.}\
  \bibnamefont {Szigeti}},\ }\bibfield  {title} {\enquote {\bibinfo {title}
  {Quantum metrology with mixed states: When recovering lost information is
  better than never losing it},}\ }\href {\doibase 10.1103/PhysRevA.92.032317}
  {\bibfield  {journal} {\bibinfo  {journal} {Phys. Rev. A}\ }\textbf {\bibinfo
  {volume} {92}},\ \bibinfo {pages} {032317} (\bibinfo {year}
  {2015})}\BibitemShut {NoStop}%
\bibitem [{\citenamefont {Ribeiro}\ \emph {et~al.}(2007)\citenamefont
  {Ribeiro}, \citenamefont {Vidal},\ and\ \citenamefont
  {Mosseri}}]{Ribeiro:2007}%
  \BibitemOpen
  \bibfield  {author} {\bibinfo {author} {\bibfnamefont {Pedro}\ \bibnamefont
  {Ribeiro}}, \bibinfo {author} {\bibfnamefont {Julien}\ \bibnamefont {Vidal}},
  \ and\ \bibinfo {author} {\bibfnamefont {R\'emy}\ \bibnamefont {Mosseri}},\
  }\bibfield  {title} {\enquote {\bibinfo {title} {Thermodynamical limit of the
  {Lipkin-Meshkov-Glick} model},}\ }\href {\doibase
  10.1103/PhysRevLett.99.050402} {\bibfield  {journal} {\bibinfo  {journal}
  {Phys. Rev. Lett.}\ }\textbf {\bibinfo {volume} {99}},\ \bibinfo {pages}
  {050402} (\bibinfo {year} {2007})}\BibitemShut {NoStop}%
\bibitem [{\citenamefont {Caneva}\ \emph {et~al.}(2012)\citenamefont {Caneva},
  \citenamefont {Calarco},\ and\ \citenamefont {Montangero}}]{Caneva:2012}%
  \BibitemOpen
  \bibfield  {author} {\bibinfo {author} {\bibfnamefont {Tommaso}\ \bibnamefont
  {Caneva}}, \bibinfo {author} {\bibfnamefont {Tommaso}\ \bibnamefont
  {Calarco}}, \ and\ \bibinfo {author} {\bibfnamefont {Simone}\ \bibnamefont
  {Montangero}},\ }\bibfield  {title} {\enquote {\bibinfo {title}
  {Entanglement-storage units},}\ }\href
  {http://stacks.iop.org/1367-2630/14/i=9/a=093041} {\bibfield  {journal}
  {\bibinfo  {journal} {New Journal of Physics}\ }\textbf {\bibinfo {volume}
  {14}},\ \bibinfo {pages} {093041} (\bibinfo {year} {2012})}\BibitemShut
  {NoStop}%
\bibitem [{\citenamefont {Pezz\'e}\ and\ \citenamefont
  {Smerzi}(2013)}]{Pezze:2013}%
  \BibitemOpen
  \bibfield  {author} {\bibinfo {author} {\bibfnamefont {Luca}\ \bibnamefont
  {Pezz\'e}}\ and\ \bibinfo {author} {\bibfnamefont {Augusto}\ \bibnamefont
  {Smerzi}},\ }\bibfield  {title} {\enquote {\bibinfo {title} {Ultrasensitive
  two-mode interferometry with single-mode number squeezing},}\ }\href
  {\doibase 10.1103/PhysRevLett.110.163604} {\bibfield  {journal} {\bibinfo
  {journal} {Phys. Rev. Lett.}\ }\textbf {\bibinfo {volume} {110}},\ \bibinfo
  {pages} {163604} (\bibinfo {year} {2013})}\BibitemShut {NoStop}%
\bibitem [{\citenamefont {Walls}\ and\ \citenamefont
  {Milburn}(2008)}]{Walls:2008}%
  \BibitemOpen
  \bibfield  {author} {\bibinfo {author} {\bibfnamefont {D.~F.}\ \bibnamefont
  {Walls}}\ and\ \bibinfo {author} {\bibfnamefont {G.~J.}\ \bibnamefont
  {Milburn}},\ }\href@noop {} {\emph {\bibinfo {title} {Quantum Optics}}},\
  \bibinfo {edition} {2nd}\ ed.\ (\bibinfo  {publisher} {Springer-Verlag},\
  \bibinfo {address} {Berlin and Heidelberg},\ \bibinfo {year}
  {2008})\BibitemShut {NoStop}%
\bibitem [{\citenamefont {Haine}(2018)}]{Haine:2017}%
  \BibitemOpen
  \bibfield  {author} {\bibinfo {author} {\bibfnamefont {S.~A.}\ \bibnamefont
  {Haine}},\ }\bibfield  {title} {\enquote {\bibinfo {title} {Quantum noise in
  bright soliton matterwave interferometry},}\ }\href
  {http://iopscience.iop.org/10.1088/1367-2630/aab47f} {\bibfield  {journal}
  {\bibinfo  {journal} {New Journal of Physics}\ } (\bibinfo {year}
  {2018})}\BibitemShut {NoStop}%
\bibitem [{not()}]{note2}%
  \BibitemOpen
  \href@noop {} {}\bibinfo {howpublished} {In order to numerically evaluate the
  optimal value of the CFI in the presence of detection noise, we have
  considered the optimization of the measurement basis in the planes normal to
  $\hat{S}_n$ as well as $\hat{S}_x$.}\BibitemShut {Stop}%
\end{thebibliography}%

\end{document}